\documentclass[11pt,a4paper]{article}
\usepackage{graphicx}
\usepackage{color}
\usepackage{amssymb}
\usepackage{amsmath}
\usepackage{multirow}
\usepackage{epsfig}
\usepackage{amscd}

\newcommand{\emailAdd}[1]{\gdef\@emailAdd{#1}}
\newcommand{\affiliation}[1]{\gdef\@affiliation{#1}}
\newcommand{\keywords}[1]{\gdef\@keywords{#1}}

\def\Journal#1#2#3#4{{#1} {\bf #2}, (#3) #4}

\def\etal{{\it et al.}}

\def\APH{\em Annals Phys.}

\def\APJ{\em ApJ.}

\def\APJS{\em ApJ.Suppl.}
\def\APP{\em Astropart. Phys.}

\def\ASP{\em ASP Conf.Ser.}

\def\CQG{\em Class.Quant.Grav.}

\def\IMA{{\em Int. J. Mod. Phys.} A}
\def\IMD{{\em Int. J. Mod. Phys.} D}

\def\JCA{\em J. Cosmol. Astrop. Phys.}

\def\JMP{\em J. Math. Phys.}

\def\JRS{\em J. Royal Stati.Soc.}

\def\MRA{\em MNRAS}

\def\NPB{{\em Nucl. Phys.} B}

\def\PLB{{\em Phys. Lett.} B}

\def\PRD{{\em Phys. Rev.} D}
\def\PRL{\em Phys. Rev. Lett.}

\def\PRE{\em Phys. Rep.}

\def\RMP{\em Rep. Mod. Phys.}
\def\RPP{\em Rept. Prog. Phys.}

\def\be{\begin{equation}}
\def\ee{\end{equation}}
\def\bea{\begin{eqnarray}}
\def\eea{\end{eqnarray}}

\begin{document}
\begin {center}
{\bf \large Discrimination between $\Lambda$CDM, quintessence, and modified gravity models 
using wide area surveys}\\
{Houri~Ziaeepour}\\
{Max Planck Institut f\"ur Extraterrestrische Physik (MPE), \\
Giessenbachstra$\mathbf{\beta}$e, Postfach 1312, 85741 Garching, Germany.}\\
{\scriptsize houriziaeepour@gmail.com}
\end{center}
\begin{abstract}
In the past decade or so observations of supernovae, Large Scale Structures (LSS), and the 
Cosmic Microwave Background (CMB) have confirmed the presence of what is called dark energy - 
the cause of accelerating expansion of the Universe. They have also measured its density as well 
as the value of other cosmological parameters according to the concordance $\Lambda$CDM model with 
few percent uncertainties. Next generation of 
surveys should allow to constrain this model with better precisions, or distinguish between a 
$\Lambda$CDM and alternative models such as modified gravity and (interacting)-quintessence 
models. In this work we parametrize both homogeneous and anisotropic components of matter density 
in the context of interacting dark energy models with the goal of discriminating between $f(R)$ 
modified gravity and its generalizations, and interacting dark energy models, for which we also 
propose a phenomenological description of energy-momentum conservation equations inspired 
by particle physics. It is based on the fact that the simplest interactions between 
particles/fields are elastic scattering and decay. The parametrization of growth rate 
proposed here is nonetheless general and can be used to constrain other interactions. As an 
example of applications, we present an order of magnitude estimation of the accuracy of the 
measurement of these parameters using Euclid and Planck surveys data, and leave a better 
estimation to a dedicated work.
\end{abstract}


\section {Introduction} \label{sec:intro}
Nowadays it is a well established fact that according to the Einstein theory of gravity 
$\sim 73\%$ of the mass and energy in the Universe is in a strange form with unusual properties 
inconsistent with any type of matter known to us. It is generally called {\it dark energy} 
(see e.g.~\cite{derev,derev1,derev2,derev3} for recent reviews). In the last two decades or so 
numerous models have been suggested to explain this mysterious and dominant constituent of the 
Universe. The majority of these theories can be classified in one the following three 
categories: 1) Models based on a scalar field e.g. quintessence~\cite{quin,quin1,quin2,quin3} 
and its variants such as K-essence~\cite{kess,kess1,kess2,kess3,kess4} in which the kinetic term 
in the Lagrangian has a non-standard form, and varying neutrino mass 
models~\cite{numassvar0,numassvar1} in which the accelerating expansion of the Universe is 
generated by the variation of neutrinos mass due to their interaction with a light scalar field; 
2) Modified gravity models in which dark energy is explained as the deviation of gravitational 
interaction from Einstein theory of gravity. Examples of such models include 
scalar-tensor~\cite{scaltens,scaltens1,scaltens2} and $\mathrm {f}(R)$ 
gravity~\cite{frgr,frgr1,frbean}, Chameleon~\cite{chamel,chamel1,chamel2}, and DGP~\cite{dgp}; 
3) A cosmological constant - introduced by Einstein himself~\cite{cc} and interpreted by 
Lema\^itre as the energy density of vacuum~\cite{vacuum}. It is phenomenologically the simplest 
of three categories, and is still the best fit to all observational 
data~\cite{sdss,wmap7,snunion}. However, naive estimations of vacuum energy are $\sim 42$ to 
$123$ orders of magnitude larger than the observed dark energy~\cite{weinbergde}. For this 
reason, alternative explanations have been explored even before the observation of the 
accelerating expansion of the Universe~\cite{quin,quin1}. The main task of cosmologists today is 
discriminating between these models, in particular distinguishing the first two categories 
mentioned above from a cosmological constant.

A notable difference between a cosmological constant and some of alternative models is the 
presence of a weak interaction between matter and dark energy. Pure quintessence 
models, in which there is no interaction between the scalar field and matter are somehow 
pathological because all known fundamental particles, including neutrinos which have very small 
couplings, interact non-gravitationally with some other particles. Even axiomatic weakly coupled 
particles such as axions~\cite{axion,axion1} are expected to interact with gauge bosons such as 
gluons. Fields in candidate extensions of the Standard Model are related to each others by 
symmetries, thus either by gauge interaction or by mass mixing. On the other hand, if a field 
such as quintessence interacts only with gravity, then naturally it should be considered to 
belong to the gravity sector. An example of such fields is dilaton which was first introduced 
in the context of Kaluza-Klein model for the unification of gravity and electromagnetic 
forces~\cite{dilaton,dilaton0}, and is also associated with conformal gravity models, see 
e.g.~\cite{grconform} and references therein. But gravity is a universal force and interacts 
with all other particles. Thus, in contradiction with the assumption above, the quintessence 
field must have an interaction with other particles. In fact dilaton does have non-minimal 
interaction with other species, see for instance~\cite{quindilaton,quindilaton0}. 
This makes the task of finding a candidate for a non-interacting quintessence field very 
difficult. A more problematic issue with pure quintessence models is the fact that they do not 
solve the coincidence problem of dark energy, i.e. why it becomes dominant only at late times and 
after galaxy formation. Interacting quintessence models in which the quintessence field has a 
weak interaction with some matter species, in particular with dark matter, can solve or at least 
soften the huge fine-tuning of dark energy density with respect to matter in the early 
Universe~\cite{deint,houriquin}.

In modified gravity models the deviation from the Einstein theory of gravity can be, either 
written explicitly, or presented by introducing new fields, usually scalars, in the matter 
sector. The first presentation is called {\it Jourdan frame} and the second {\it Einstein 
frame}. Because in the latter case the model looks very similar to an interacting quintessence 
model\footnote{Note that when we talk about interacting quintessence models we mean models in 
which the scalar field interact with some other constituents of the Universe. All quintessence 
models have a self-interaction which is not explicitly considered in the formulation presented 
in this work}, it is necessary to find a proper definition that discriminates between what is 
called {\it modified gravity} and what is called {\it interacting quintessence}. In 
modified gravity, the 
scalar field is usually related to a dilaton field, thus it has a geometrical origin and arises 
from a broken conformal symmetry~\cite{scaltens}. For this reason, the scalar field always 
interacts with the trace of energy-momentum tensor of matter~\cite{frbean}. The situation is not 
so straightforward for interaction between matter and scalar field in interacting quintessence 
models for which various types of interactions are considered in the literature, see for 
instance~\cite{quinintterm}. In many of these models in analogy with modified gravity, in 
particular $\mathrm {f}(R)$ models, the interaction term is considered to be proportional to the 
trace of the energy-momentum tensor of matter\footnote{For the reasons described in detail in 
Sec. \ref{sec:interact}, when we talk about the interaction term, we mean the modification of 
energy-momentum conservation equation due to an interaction.}. 

In this work we try to determine the interaction between dark matter and dark energy in a 
collisional description of interactions inspired by particle physics. Using the Boltzmann 
equation with a collisional term and some results from studies of the microphysics of dark 
energy condensate~\cite{houricondens}, we show that the interaction can be described only 
approximately by spacetime dependent functions, and in general one needs the distribution in 
the phase space $f(x,p)$ where $p$ is the 4-momentum, see Sec. \ref{sec:interact} for more 
details. However, at present and for foreseeable future, we cannot observe the phase space 
distribution of dark energy. Considering this fact, we use thermodynamical description of 
average energy-momentum and velocity to obtain approximate covariant expressions for interactions 
between matter and a scalar field as dark energy. This leads to a modification of 
energy-momentum conservation equation which explicitly deviates from modified gravity. Their 
difference can be used as a mean for classifying models and discriminating these two categories 
in the data.

On the observational side, one has to find the best way of parametrizing cosmological evolution 
equations such that they admit discrimination between at least the three major categories of 
models discussed above. In preference they should not depend on the details which are neither 
well understood nor can be targeted with the precision of present and near future surveys. 
Observations show that dark energy has negligible clustering (see 
e.g.~\cite{declustering0,declustering,declustering1,declustering2} for latest results). Therefore, 
its dominant contribution is in the homogeneous component of the Einstein and conservation 
equations. It also affects the evolution of anisotropies mainly through their dependence on the 
background cosmology. For this reason, irrespective of the way we measure the equation of state 
of dark energy - from observations of supernovae that are only sensitive to background cosmology, 
or from observations of matter perturbations by measuring lensing or galaxy distribution - we 
must extract parameters of background cosmology to determine the contribution of dark energy. 
Consequently, it is crucial to understand how different models affect this component through a 
proper parametrization that facilitates the discrimination between various models. This is 
another goal of the present work. Although there are few popular 
parametrizations~\cite{param0,param1,param2,param3} in the literature, specially for testing 
modification of the Einstein theory of gravity at large scales, as we will show in this work, 
they are not suitable for discriminating between modified gravity and (interacting)-quintessence 
models. We should remind that for $\Lambda$CDM model the growth rate $\mathbf {f}$ is roughly 
scale independent. Therefore, observation of the violation of this property would be a clear 
signature of inconsistency with standard cosmology. But the measurement of $\mathbf {f}$ and 
the expansion rate $H$ by themselves are not enough for discriminating between modified gravity, 
quintessence, and interacting quintessence, and a parametrization that does not depend on the 
details of these models is necessary to highlight their differences. Evidently, one can simply 
fit the data with models and compare their goodness of fit. But, this does not take into 
account the degeneracies and similarities. Therefore, a smart parametrization and better data 
analysing methods are necessary. Moreover, the fact that most popular modified gravity models 
can be formulated as an scalar field theory means that their differences from 
(interacting)-quintessence must be understood and the parametrization must be performed in a 
way that it highlights these differences and help discrimination.

In this work we propose a new set of parameters to describe, in a model independent way, the 
effect of an interacting dark energy on the evolution of the expansion rate of the Universe and 
another set of parameters for the growth rate. These quantities are the most sensitive 
measurables for discriminating between dark energy models. Consequently, the ultimate goal of 
various measurement methods is to constrain cosmological and dark energy models by measuring one 
or both these quantities. For instance, galaxy distribution and lensing surveys determine the 
power spectrum of fluctuations for one or multiple redshift bins. Future large surface and 
sensitive spectroscopic surveys such as Euclid allow to determine the matter power spectrum for a 
statistically significant number of redshift bands, and thereby extract the growth rate, see e.g.
~\cite {wigglezgr} for the methodology applied to The WiggleZ Dark Energy Survey. The BAO 
measurements determine the expansion rate and angular diameter distance at one or multiple 
redshift bins, and can thereby estimate variation of these quantities. Supernovae data measure 
the expansion rate directly, and the variation of $H(z)$ can be extracted. Therefore, 
parametrizations that we will discuss here are relevant for all measurement methods in cosmology.

In Sec. \ref{sec:frieman} we present a new parametrization for Friedman equation in the context 
of a general interacting dark energy model\footnote{After the submission of this paper a similar 
parametrization for the Friedman equation is reported independently by~\cite{ffparam}.}. In 
Ref.~\cite{houriaz} we defined a quantity $B (z) \propto d\bar{\rho}/dz$ and proposed 
it for the measurement of the equation of state of dark energy defined as $w \equiv P_{de}/
\rho_{de}$ where $P_{de}$ and $\rho_{de}$ are pressure and energy density of dark energy 
respectively. It is specially suitable for measuring the deviation from a cosmological constant, 
see Appendix \ref{app:a} for the definition of $B (z)$ and a review of its properties. 
In addition, we argued that in what concerns the sign of $\gamma (z)$ (see equation 
(\ref{gammade}) below for its definition), this quantity has distinct geometrical properties 
which make it less sensitive to uncertainties of other quantities such as $H_0$ or $\Omega_m$, 
respectively the present value of Hubble constant and the density fraction of matter. The sign of 
$\gamma (z)$ is the discriminator between what is called phantom models which have $w < -1$, and 
normal scalar fields (quintessence) models and a cosmological constant for which $w \geqslant -1$.

Using this parametrization and properties of $B (z)$, we show that in presence of an interaction 
between dark energy and other components, one obtains different estimation for $\gamma_{de}^{eff}$ 
(see next section for its definition) from $H(z)$ and from $B(z)$ when data is analyzed with the 
null hypothesis of a $\Lambda$CDM model as dark energy. In this way one can predict the 
sensitivity of surveys to interacting dark energy models in a model-independent manner. Then, 
we discuss the properties of parameters for each category of models, their differences, 
and how this information can be used to discriminate between various dark energy models. In 
Sec. \ref{sec:interact} we describe phenomenological interactions for interacting quintessence 
models and compare it with modified gravity. This leads to an approximate description for 
non-gravitational interactions between dark matter and dark energy. 

In Sec. \ref{sec:perturb} we present evolution equations of over-density and velocity fields in 
each category of models for the interactions obtained in Sec. \ref{sec:interact}. Then we describe 
how one can discriminate between interacting quintessence and modified gravity models by using 
matter power spectrum and its evolution, i.e. the growth rate of anisotropies. Because the growth 
rate plays a special role in the discrimination between various dark energy models, in 
Sec. \ref{sec:forecast} we parametrize its evolution, and as an example of application, we 
obtain an order of magnitude estimate for the discrimination ability of the Euclid 
mission~\cite{euclid} that measures both parameters of the homogeneous component (the background 
cosmology) and the evolution of growth rate of matter anisotropies. In addition, we compare our 
parametrization with other parametrizations that can be found in the literature which are usually 
designed to test the Einstein theory of gravity. Conclusions and outlines are summarized in 
Sec. \ref {sec:outline}. Properties of the functions $B (z)$ (and $A(z)$) are reviewed in 
Appendix \ref{app:a}. Fisher matrix for dark energy without parametrization of its equation of 
state $w(z)$ is described in Appendix \ref{app:b}. A summary of covariant formulation of a 
classical scalar fields as a perfect fluid is given in Appendix \ref{app:c}. In Appendix 
\ref{app:d} we calculate an approximate analytical solution for the growth rate of matter 
anisotropies.

Here we must emphasize that the predictions for future missions obtained in this work are only 
representative and order of magnitude estimations of what is expect from future surveys. They 
should be considered as a {\it QD} (Quick and Dirty), hand-shaking predictions. Their purpose 
is only to show that it is possible to measure the new parameters with reasonable uncertainties. 
A proper prediction for future observation projects needs detailed consideration of instrumental 
response, simulation of data analysing procedure, and understanding of the sources of systematic 
and statistical errors. Fulfilling these requests necessitates a dedicated investigation which is 
out of the scope of the present work that targets theoretical issues related to the 
discrimination between various dark energy models. In fact, a number of authors have performed 
predictions for uncertainties of various measured quantities by future missions, see for 
instance~\cite{frsilvestri,predictions,predictions1,predictions2}. They usually consider models 
that can be classified as modified gravity according to the classification criteria discussed 
in Sec. \ref{sec:interact}. Nonetheless, some of their parameters can be related to quantities 
defined in this work, thus their predictions can be used to obtain a rough estimation for the 
expected uncertainties for the new parameters. 

Throughout this work we use Einstein frame for modified gravity models unless it is explicitly 
specified. In this way, a unified description can be made for all interacting dark energy models 
based on a scalar field formulation.

\section{Friedman equation in interacting dark matter-dark energy models} 
\label{sec:frieman}
Apriori the measurement of the equation of state of dark energy is simple. It is enough to 
measure the expansion rate of the Universe $H(z) \equiv \dot {a}(z)/a$, or a quantity related 
to it such as the luminosity distance $D(z)$ at different 
redshifts. Then, by modeling known constituents of the Universe as non-interacting perfect 
fluids, one can fit the data and measure the effective equation of state of dark energy 
$w_{eff}(z)$, defined as $P_{eff}(z)/\rho_{eff}(z)$. The suffix {\it ``eff''} is used to remind 
that pressures and densities obtained in this way can be effective quantities rather than physical 
pressure and density of constituents, because we have neglected any interaction between 
components. Therefore, from now on {\it effective quantities} mean quantities determined 
from data by considering a null hypothesis.

In practice however the life is not so simple. The density of a perfect fluid changes with 
redshift as $(1+z)^{3\gamma}$ ($\gamma$ is defined in (\ref{gammade})). Therefore, at low 
redshifts when $z \rightarrow 0$, the total density is not very sensitive to the value of 
$\gamma$ or equivalently $w(z)$ and their variation with $z$, see Appendix \ref{app:a} for more 
details. This statement is independent of the type of data or proxy used for determining 
$H(z)$ or $D(z)$. On the other hand, at high redshifts where $H(z)$ is more sensitive to the 
equation of state, dark energy is subdominant. Moreover, it is more difficult to measure $H(z)$ 
and $D(z)$ at higher redshifts and measurement uncertainties can make the estimation of $w(z)$ 
and its evolution unusable for discrimination between models.

If constituents of the Universe do not interact with each other, Friedman equation 
which determines the evolution of expansion function $a(t)$ can be written as:
\bea
&&\frac{H^2}{H_0^2} = \frac {\rho (z)}{\rho_0} = \Omega_m (1+z)^3 + \Omega_h (1+z)^4 + 
\Omega_K (1+z)^2 + \Omega_{de} (1+z)^{3\gamma(z)}, \quad {\rho_c (z)} \equiv \frac{3H^2}{8\pi G} 
\label{friedmannoint} \\
&&\gamma (z) = \frac{1}{\ln (1+z)} \int_0^z dz'\frac{1 + w(z')}{1+z'}, \quad 
P_{de} (z) \equiv w(z) \rho_{de} \label{gammade} \\
&& \text{\it m = cold dark matter, b = baryons, h = hot matter, k = curvature, and 
de = dark energy} \nonumber
\eea
In this class of models matter and radiation densities evolve only due to the expansion. This is 
a good approximation for all redshifts $z < z_{cmb} \sim 1100$.

In interacting dark energy models matter and radiation terms in the right hand side of the 
Friedman equation (\ref{friedmannoint}) can contain an additional redshift-dependent factor:
\be
\frac{H^2}{H_0^2} = \frac {\rho_c (z)}{\rho_{c0}} = \sum_i \Omega_i {\mathcal F}_{i}(z)
(1+z)^{3\gamma_i} \quad \quad i=\text{{\it m, b, h,k,} and {\it de}} \label{friedmanintde}
\ee
Without lack of generality we assume that ${\mathcal F}_{de} = 1$ and all redshift dependent 
terms are included in $\gamma (z)$. In quintessence models the coefficient of the curvature 
term also is constant because it is assumed to be related to geometry/gravity and independent 
of the behaviour of other components. At present observations are consistent with only 
gravitational interaction between various components in (\ref{friedmanintde}), thus additional 
interactions must be very weak. By definition and without lack of generality we consider 
${\mathcal F}_i(z=0) = 1$. Observations also show that $\Omega_k \approx 0$, therefore 
throughout this work we assume $\Omega_k = 0$ unless it is explicitly mentioned. Note that in 
the case of modified gravity models, a parametrization similar to (\ref {friedmanintde}) can be 
defined both in Einstein and Jordan frames.

A simple example for which an approximate expression for ${\mathcal F}_i(z)$ coefficients can 
be found is a model with a cosmological constant as dark energy and a slowly decaying dark 
matter. The decay remnants are assumed to be visible relativistic particles~\cite{houridecay}. 
In this case:
\bea
&& \frac{H^2}{H_0^2} \approx {\Omega}_m (1 + z)^3 \exp (\frac {{\tau}_0 - t}{\tau}) + 
{\Omega}_b (1 + z)^3 + {\Omega}_h (1 + z)^4 + {\Omega}_m (1 + z)^4 \biggl (1 - 
\exp (\frac {{\tau}_0 - t}{\tau}) \biggr ) + {\Omega}_{\Lambda} \nonumber \\
&& \label {totdens} \\
&& {\mathcal F}_m(t) \approx \exp (\frac {\tau_0 - t}{\tau}) + (1+z) \biggl (1 - \exp 
(\frac {t_0 - t}{\tau}) \biggr ), \quad \tau \gg \tau_0, \quad {\mathcal F}_b = {\mathcal F}_h = 1
\quad \gamma (z) = 0 \label{dmdecaycoeff}
\eea
where $\tau$ is the lifetime of dark matter and $\tau_0$ is the age of the Universe. It is 
demonstrated that in this example, if the decay/interaction of dark matter is not considered in 
the data analysis, a $w_{eff} < -1$ can be obtained for dark energy, see~\cite{houriustat,das05} 
for more details about the set up and the proof. 

Note that in (\ref{dmdecaycoeff}), we have included the contribution of relativistic remnants 
in ${\mathcal F}_m$. However, as this component has a redshift dependence similar to hot matter, 
it also makes sense to consider it as hot matter and add it to hot component. It is even 
possible to add this term to dark energy contribution, as long as it is small and induces only 
a slight deviation from a cosmological constant. In this case, one can show that 
the {\it effective dark energy} will have $w_{eff} < -1$~\cite{houriustat}. The reason for such 
freedom is that we do not measure or take into account the decay remnants. This example 
clearly shows that parametrization (\ref{friedmanintde}) is not unique when all the components 
and their interactions are not know. Therefore, one has to be very careful about 
degeneracies when data are analyzed and interpreted. In particular, prior assumptions such as 
stability of matter and radiation components can affect measurements and conclusions. This 
example also show that for ruling out $\Lambda$CDM model, it is enough to prove that at least 
one of ${\mathcal F}_i \neq 1$, or $\gamma_{de} \neq 0$\footnote{This statement is true if baryon 
pressure is negligible. Future surveys can be sensitive to small baryon pressure. In this case 
it must be taken into account before any conclusion about $\Lambda$CDM model is made.}.

Extension of this example to quintessence models without coupling to matter is straightforward 
and one simply needs to consider $\gamma (z) \neq 0$. A more interesting extension is to assume 
that the quintessence scalar field is one of the remnants of the decay of dark matter, which 
during cosmological time condensates and makes a classical quintessence field. In this case, 
it has been shown~\cite{houriquin,houriquin1,houricondens} that coefficients ${\mathcal F}_m$, 
${\mathcal F}_h$, and equation of state of dark energy $w(z)$ (or equivalently $\gamma (z)$) 
are not independent. However, their relations are too sophisticated and cannot be described in an 
analytical form and numerical techniques should be employed~\cite {houriquin}.

According to (\ref{dmdecaycoeff}):
\be
{\mathcal F}_m(z) > {\mathcal F}_m(z=0) \label{quinfdmtode}
\ee
and because $\tau \gg \tau_0$, ${\mathcal F}_i(z)$ coefficients are close to 1 at all redshifts. 
In general, for an interaction which transfers energy from dark matter to other components, the 
inequality (\ref{quinfdmtode}) is applied because at high redshifts one expects a larger 
contribution of dark matter in the total density than in a non-interacting model. Inversely, if 
energy is transferred from other components, for instance from dark energy, to dark matter:
\be
{\mathcal F}_m(z) < {\mathcal F}_m(z=0) \label{quinfdetodm}
\ee
An example of such models is {\it scaling dark energy}~\cite{scalingde,scalingde1} in which 
at early times dark energy has a much larger contribution in the total energy density, but it 
gradually decays to dark matter and only recently its equation of state approaches $w \sim -1$. 
Another example is the class of models called {\it early dark energy}. Although the original
model~\cite{earlyde,earlyde1,earlyde2} is a pure quintessence/k-essence, there are variants of 
this model in which, there is an interaction in the dark sector~\cite{earlydeint} or between 
dark energy and visible sector~\cite{earlydeintvarpalpha}.

In models with elastic interaction between two sectors, no energy is transferred between them, and 
${\mathcal F}_m(z) = {\mathcal F}_h(z) = 1$. Nonetheless, the phase space of matter and dark 
energy in these models can change and thereby $w_{de}$ can depend on $z$. 

For $\mathrm {f}(R)$ modified gravity models homogeneous Einstein equations and energy 
conservation equation in Jordan frame are~\cite{frbean}:\footnote{When equations apply to both 
dark matter and baryons, we indicate them collectively with subscript $m$.}
\bea
&& (1+\mathrm {f}_R) H^2 + \frac{1}{6} \mathrm {f} - \frac{a''}{a^3}\mathrm {f}_R + 
\frac{\mathrm {f'}_R}{a} H = \frac {8\pi G}{3} \sum_i \rho_i \label{grmg0} \\
&& \frac {a''}{a^3} = -\frac{4\pi G}{3} \sum_i (\rho_i + 3P_i) + (1+\mathrm {f}_R) H^2 + 
\frac{\mathrm {f}}{6} - 
\frac {H\mathrm {f}'_R}{a} - \frac{\mathrm {f}''_R}{2a^2} \label{grmg1} \\
&& \dot{\rho}_i + 3H\rho_i = -\frac{\dot{f}_R}{2 (1+\mathrm {f}_R)}(\rho_i -3 P_i), 
\quad i=\text{\it m, h, k} \label{grgmener}
\eea
where $a' = a\dot{a}$.\footnote{Here we have written Einstein and conservation equations in 
Jordan frame because they lead to expressions for ${\mathcal F}_i$ coefficients which are 
explicitly very different from quintessence case.} Dot and prime mean derivation with respect to 
comoving and conformal time, respectively. Subscript $R$ means derivation with respect 
to scalar curvature $R \equiv R_{\mu\nu}g^{\mu\nu}$. We remind that at linear order the effect of 
matter perturbations on $R$ is zero, thus $R$ only depends on $z$ and the effect of 
$\mathrm {f}(R) \neq 0$ on the evolution of perturbations manifests itself by changing the 
background cosmology. 

After solving density conservation equation (\ref{grgmener}), Friedman equation (\ref{grmg0}) 
can be written as the following:
\bea
\rho_i (z) &=& \rho_i (z=0) (1+z)^{3\gamma_i} \biggl (\frac{{1+\mathrm {f}_R (z=0)}}
{1+\mathrm {f}_R (z)} \biggr)^{-\frac{1-3w_i}{2}} \label{rhosolmg} \\
\frac{H^2}{H_0^2} &=& \frac {\rho_c (z)}{\rho_{c0}} = \sum_i \Omega_i {\mathcal F}_{i}(z)
(1+z)^{3\gamma_i}  \label{friedmanmg} \\
\rho_{de} &=& \frac{3}{8\pi G}~\frac{1}{1+\mathrm {f}_R}(-\frac{\mathrm {f}(R)}{6} + 
\frac{a''}{a^3}\mathrm {f}_R - H\dot{\mathrm {f}}_R) \label{mgdedens} \\
{\mathcal F}_i(z) &=& \biggl (\frac{1+\mathrm {f}_R(R(z=0))}{1+\mathrm {f}_R(R(z))}\biggr)^
{-\frac{1-3w_i}{2}}, \quad w_m = 0, \quad w_h = \frac{1}{3}, \quad w_k = -\frac{1}{3}
\label{fmgdef}
\eea
Equation (\ref{mgdedens}) is the energy density of {\it effective dark energy} in 
$\mathrm {f}(R)$ gravity models. Similar to quintessence models we can assume 
${\mathcal F}_{de} = 1$. The only explicit difference between (\ref{friedmanmg}) and the same 
equation for an interacting quintessence model is the presence of a nontrivial coefficient for 
the curvature term if $\Omega_k \neq 0$. Nonetheless, the evolution of coefficients 
${\mathcal F}_i(z)$ with redshift is different from their counterparts in interacting 
quintessence models, in particular from models in which energy is transferred to dark energy at 
low redshifts, see equation (\ref{quinfdmtode}). In fact, the function $\mathrm {f}(R)$ is not 
completely arbitrary and must satisfy a number of constraints. Notably, 
$\mathrm {f}(R)|_{|R| \gg 0} \rightarrow 0$ to make the model consistent with Einstein theory of 
gravity in mild or strong gravity fields, and $\mathrm {f}_R > 0$ due to stability 
constraint~\cite{modgrstabil}. Under these conditions:
\be
{\mathcal F}_i(z) > {\mathcal F}_i(z=0)  \label{mgfz}
\ee
Comparing (\ref{mgfz}), (\ref{quinfdmtode}), and (\ref{quinfdetodm}) one can immediately 
conclude that the measurement of ${\mathcal F}_m(z)$ and its evolution with redshift can 
discriminate between dark energy models in which energy is transferred from dark energy to 
dark matter such as scaling models, and $\mathrm {f}_R$ modified gravity models. But it cannot 
discriminate modified gravity from models in which energy is transferred from dark matter to 
dark energy such as the model discussed in~\cite{houriquin,houriquin1,houricondens}. To 
discriminate the latter and other models of this category from $\mathrm {f}_R$ modified 
gravity, the coefficient of relativistic (hot) component ${\mathcal F}_h(z)$ and its evolution 
must be measured. Evidently, such measurements are very difficult. For instance, one has to 
measure very precisely the temperature of CMB at high redshifts or $H(z)$ at a large number of 
redshift bins and fit the data with ${\mathcal F}_h \neq 1$. In Einstein frame the evolution of 
matter density is the same as in equation (\ref{rhosolmg})~\cite{frbean}, but evolution equation 
of hot matter is similar to $\Lambda$CDM. In what concerns the discrimination from interacting 
quintessence what is discussed from Jourdan is applicable.

\subsection {Model-independent discrimination of interacting dark energy models} 
\label{sec:discrim}
In this section we show that if $\Lambda$CDM or a simple quintessence are considered as null 
hypothesis, measurements of effective dark energy density and effective equation of state from 
$H (z)$ and the function $A(z)$ defined in Appendix \ref {app:a} separately, give different 
values for these quantities if dark energy interacts with matter. Similarity of 
${\mathcal F}_m(z)$, specially if the curvature of the Universe is zero, means that we cannot 
distinguish between interacting quintessence and modified gravity models in a model-independent 
manner - except for the cases explained above. For this reason in this section we only study 
the discrimination between interacting dark energy models parametrized as in equation 
(\ref{friedmanmg}) and a cosmological constant and/or non-interacting quintessence.

For analyzing cosmological data, $\Lambda$CDM with a stable and non-interacting dark matter is 
usually used as null hypothesis. Nonetheless, the methodology explained below is not sensitive 
to redshift dependence of $\gamma_{de}$, and we can consider the more general case of 
non-interacting quintessence as the null hypothesis. The expansion of the Universe for 
such cosmologies is ruled by equation (\ref{friedmannoint}). Therefore, we rearrange terms in 
equation (\ref{friedmanmg}) such that it looks similar to equation (\ref{friedmannoint}). 
Then, we determine effective quantities which are measured by fitting a $\Lambda$CDM or a 
non-interacting quintessence model to data:
\be
\frac{H^2}{H_0^2} = \sum_i \Omega_i (1+z)^{3\gamma_i} + \sum_i \Omega_i ({\mathcal F}_{i}(z) - 
1) (1+z)^{3\gamma_i} + \Omega_{de} (1+z)^{3\gamma_{de} (z)} \label{friedmanmgequiv}
\ee
In null hypothesis model only $\gamma_{de}$ is redshift dependent and $\gamma_i,~i=m,~h,~k$ are 
constant. By comparing (\ref{friedmanmgequiv}) with (\ref{friedmannoint}) the {\it effective} 
contribution of dark energy is expressed as:
\be
\Omega_{eff}^{(H)} (1+z)^{3\gamma_{eff}^{(H)}(z)} = \sum_i \Omega_i ({\mathcal F}_i(z) - 1) 
(1+z)^{3\gamma_i} + \Omega_{de} (1+z)^{3\gamma_{de} (z)} \label{deeffdef}
\ee
In both interacting quintessence and modified gravity models coefficients $F_i$'s are defined 
such that $F_i (z=0) = 1$, therefore at $z = 0$ the first term in (\ref{deeffdef}) is null and 
we can separate $\Omega_{eff}$ and $\gamma_{de} (z)$:
\bea
&& \Omega_{eff}^{(H)} = \Omega_{de}, \quad \quad \gamma_{eff}^{(H)}(z=0) = \gamma_{de}(z=0) 
\label{omegaffh} \\
&& \gamma_{eff}^{(H)}(z) = \frac{\log \biggl (\sum_i \frac{\Omega_i}{\Omega_{de}}
({\mathcal F}_i (z) -1)(1+z)^{3\gamma_i} + (1+z)^{3\gamma_{de} (z)}\biggr )}{3 \log(1+z)} 
\label{deeffexp}
\eea
where superscript $(H)$ means measured from Hubble constant $H$.
 
Suppose we can also measure $A(z)$ defined in (\ref{azdef}), and use it to determine the effective 
density and equation of state of dark energy. For an interacting dark energy model parametrized 
according to (\ref{friedmanmgequiv}) quantities $B(z)$ and $A(z)$ are:
\bea
B(z) &\equiv& \frac{1}{3 (1+z)^2 \rho_0} \frac {d\rho}{dz} = \nonumber \\
&& \sum_{i=m,h,k} \Omega_i \biggl (\gamma_i {\mathcal F}_i (z) + (1+z) 
\frac{d{\mathcal F}_i}{dz} \biggr )(1+z)^{3 (\gamma_i - 1)} + \Omega_{de}(w(z)+1)
(1+z)^{3(\gamma_{de}(z) - 1)} \label{bzintdmde} \\
A(z) &\equiv& B(z) - \sum_{i=m,h,k} \Omega_i \gamma_i (1+z)^{3(\gamma_i - 1)} = \nonumber \\
&& \sum_{i=m,h,k} \Omega_i \biggl (\gamma_i ({\mathcal F}_i (z) - 1) + (1+z) 
\frac{d{\mathcal F}_i}{dz} \biggr ) (1+z)^{3 (\gamma_i - 1)} + \Omega_{de}(w(z)+1)
(1+z)^{3(\gamma_{de}(z) - 1)} \label{azintdmde}
\eea
Using (\ref{azdef}) in Appendix \ref{app:a} as the definition of $A(z)$, we find the following expression for its parameters:
\bea
\Omega_{eff}^{(A)} (w_{eff}^{(A)}(z) + 1) (1+z)^{3\gamma_{eff}^{(A)}(z)} &=& \sum_i \Omega_i 
\biggl (\gamma_i ({\mathcal F}_i (z) - 1) + (1+z) \frac{d{\mathcal F}_i}{dz} \biggr ) 
(1+z)^{3\gamma_i} + \nonumber \\
&& \Omega_{de}(w(z)+1)(1+z)^{3\gamma_{de}(z)} = (1+z) A(z) \label{deeffdefaz}
\eea
where superscript $(A)$ means measured from $A(z)$. Equations (\ref{deeffdef}) and 
(\ref{deeffdefaz}) are fundamentally different. In particular:
\be 
\Omega_{eff}^{(A)} = \frac{\sum_i \Omega_i \frac{d{\mathcal F}_i (z=0)}{dz} + \Omega_{de} 
(w(z=0)+1)}{w_{eff}^{(A)}(z=0) + 1} \label{omegaedeffaz}
\ee
which in contrast to $\Omega_{eff}^{(H)}$, in general is not equal to $\Omega_{de}$. Equality arises 
only when ${\mathcal F}_i$ do not vary with redshift i.e. ${\mathcal F}_i = 1$ at all redshifts. 
This condition is satisfied by the null hypothesis $\Lambda$CDM and by non-interacting 
quintessence models. Therefore, assuming that $\Omega_m$ and $\Omega_k$ are known (e.g. from 
CMB), simultaneous measurements of $H(z)$ and $A(z)$ at even one $z > 0$ is apriori enough for 
testing the presence of an interaction between dark matter and dark energy independent of the 
underlying model. Evidently, in practice the measurements must be performed at many redshift 
bins to improve statistics and to compensate for measurement errors.

Apriori one can use other quantities such as angular diameter distance $D_A$ or luminosity 
distance $D_L$ which are easier to measure rather than $A(z)$. However, both these quantities 
are functional of $H(z)$ - 
through integration of $1/H^{1/2}(z)$. Thus, in general they do not have an analytical expression. 
Besides, their derivatives depend on ${\mathcal F}_i$'s only, in contrast to (\ref{deeffdefaz}) 
that depends on both ${\mathcal F}_i$'s and their derivatives. Therefore, $\Omega_{eff}$ and 
$\gamma_{eff}$ obtained from $dD_A/dz$ or $dD_L/dz$ will be equal to ones determined from $H(z)$ 
irrespective of the underlying cosmology. This shows that the function $A(z)$ 
(or equivalently $B(z)$) introduced in~\cite{houriaz} has special properties and is well suited 
for discriminating between dark energy models. It can be measured from supernovae data, 
see~\cite{houriaz} for the methodology. As for LSS data, one needs to determine both $H(z)$ 
and its evolution $dH(z)/dz$ to be able to calculate $A(z)$, for instance from the BAO and the 
power spectrum of matter fluctuations~\cite{psobs}. This is not an easy task. As an example 
consider supernovae observations that measure the luminosity distance $D_L$ to a supernova from 
its standardized apparent magnitude. The angular luminosity distance $D_A$ is related to the 
luminosity distance, see (\ref{dah}). To determine $dD_A/dz$ apriori one can use the measured 
$D_A$, and determine its derivative (slope). However, due to scattering and discreteness of data, 
such a measurement will have large uncertainties. The same problem arises for $dH(z)/dz$ or 
$A(z)$ because they depend on derivatives of $D_L$, see equations (\ref{rhoderdl}) and 
(\ref{rhoderda}). 
Nonetheless, there are various methods such as binning of data, using a fit in place of discrete 
data, etc. that allow to improve the estimation. Near future large area surveys such as 
Euclid~\cite{euclid}, BigBOSS~\cite{bigboss}, LSST~\cite{lsst} will be able to determine these 
quantities with relatively good precision, see also Sec. \ref{sec:forecast} for measurement 
methodology. In particular, large surface spectroscopic and lensing surveys such as Euclid are 
able to determine the variation of total density with redshift $d\rho/dz \propto B(z)$ with good 
precision. In Appendix \ref{app:b} we obtain the Fisher matrix for dark energy parameters without 
considering a specific parametrization for the equation of state $w(z)$.

\subsection{Discrimination precision} \label{sec:errordiscrim}
Measurements of cosmological parameters show that $w^{obs}_{de} \sim -1$ irrespective of which 
proxy or measurement method - supernovae, CMB, or LSS has been used. This means that 
$|{\mathcal F}_i(z) - 1| \approx 0$ and $d{\mathcal F}_i (z)/dz \approx 0$. Moreover, addition 
of ${\mathcal F}_i(z)$ to the model increases the number of parameters. 
Giving the fact that we have essentially two observables: $H(z)$ and one of $D_A(z)$, $D_L(z)$ 
or $B(z)$, greater number of parameters means also greater degeneracy, thus more uncertainty 
for discrimination between $\Lambda$CDM, a non-interacting quintessence, and interacting dark 
energy models. 

One way of measuring the presence of interaction without having to fit data to the large number 
of parameters in equations (\ref{friedmanmgequiv}) and (\ref{azintdmde}), is to measure how 
different $\Omega_{eff}^{(H)},~\Omega_{eff}^{(A)},~\gamma_{eff}^{(H)},$ and $\gamma_{eff}^{(A)}(z)$ are, 
because as we discussed in the previous section, when ${\mathcal F}_i \neq 1$ these effective 
quantities are not the same. To this end, a natural criteria is:
\be
\Theta (z) \equiv \frac{\Omega_{eff}^{(A)} (w_{eff}^{(A)}(z) + 1) 
(1+z)^{3\gamma_{eff}^{(A)}(z)} - \Omega_{eff}^{(H)} (w_{eff}^{(H)}(z) + 1) 
(1+z)^{3\gamma_{eff}^{(H)}(z)}}{\Omega_{eff}^{(H)} (w_{eff}^{(H)}(z) + 1) 
(1+z)^{3\gamma_{eff}^{(H)}(z)}} \label{deftheta}
\ee
This quantity can be explained explicitly as a function of $\Omega_i,~{\mathcal F}_i,~\gamma_i$, 
and is zero when $F_i= 1,~dF_i/dz = 0$. Note that we have chosen expression 
(\ref{deeffdefaz}) for comparison rather than (\ref{deeffdef}) because it is not possible to 
determine $\Omega_{eff}^{(A)}$ in a model independent manner, see equation (\ref{omegaedeffaz}). 
By contrast $\Omega_{eff}^{(H)} = \Omega_{de}$, thus $\gamma_{eff}^{(H)}$ and thereby $w_{eff}^{(H)}$ 
can be determined without any reference to ${\mathcal F}_i$ coefficients. In~\cite {houriaz} 
we suggested to use the sign and evolution of $A(z)$ to discriminate between dark energy with 
$\gamma (z) \neq 0$ and a cosmological constant. Here $\Theta (z)$ plays a similar role for 
discriminating between interacting or non-interacting dark energy. 

Assuming that $\Omega_m$ and $\Omega_h$ are determined independently and with very good 
precision, for instance from CMB anisotropies with marginalization over $\gamma_{de}$, $\Theta$ 
can be determined from the measurement of $H(z)$ and $B(z)$. The latter can be measured from 
whole sky or wide area spectroscopic surveys data such as Euclid, or multi-band photometric 
surveys such as DES. Evidently determination of $B(z)$ that depends on $dH/dz$ is very difficult. 
However, it is easy to see that there is no other quantity which can be measured more easily and 
discriminates between $\Lambda$CDM and dynamical dark energy models with a better precision. 
For instance, the BAO method determines $H(z)$ and $D_A(z)$ directly. But, $D_A(z)$ depends on 
$w(z)$ or equivalently $\gamma (z)$ through an integral, see equation (\ref{dah}). Therefore, it 
is less sensitive to the variation of $\gamma (z)$ with redshift. This is analogue to binning a 
data. Evidently, a binned data is less noisy and has a smaller uncertainty. But, if the goal is 
to measure the variation of data, the binning can completely smear out small variations. 
Therefore, irrespective of methods and measured proxies, we are limited by inherent properties 
of the physical system. In this respect, the precision with which $\Theta (z)$ can be measured 
gives the ultimate sensitivity of an observation/data set to deviation from $\Lambda$CDM.

\section{Interactions} \label{sec:interact}
In the previous section we used Friedman equation for parametrizing interaction between matter 
and dark energy. Evolution of their densities is ruled by energy-momentum conservation. But, 
in presence of non-gravitational interactions between constituents the energy-momentum tensor 
of each component $T^{\mu\nu}_i$ is not separately conserved, and conservation equation can be 
only written for the total energy-momentum tensor $T^{\mu\nu}$ defined as:
\bea
&& T^{\mu\nu} \equiv \sum_i T^{\mu\nu}_{i(free)} + T^{\mu\nu}_{int} \label{ttotdef} \\
&& T^{\mu\nu}_{;\nu} = \sum_i T^{\mu\nu}_{i(free)~;\nu} + T^{\mu\nu}_{int~;\nu} = 0 
\label{ttotcons}
\eea
where $T^{\mu\nu}_{i(free)}$ is the energy-momentum tensor of component $i$ in absence of 
interaction with other components, i.e. $T^{\mu\nu}_{i(free);\nu} = 0$, and $T^{\mu\nu}_{int}$ is the 
energy-momentum tensor of interaction\footnote{Non-gravitational interactions between 
cosmological constituents must be weak. Therefore, separation of energy-momentum tensor to free 
and interaction component is allowed.}, and $T^{\mu\nu}_{int;\nu} = 0$. In the literature on 
interacting dark energy models (see e.g. ~\cite{deint}) when only two constituents - matter and 
dark energy - are considered, the energy-momentum conservation equations are usually written as:
\be
T^{\mu\nu}_{m~;\nu} = Q^\mu, \quad T^{\mu\nu}_{\varphi~;\nu} = -Q^\mu \label{tqcons}
\ee
for an interaction current $Q^\mu$. Comparing (\ref{ttotcons}) and with (\ref{tqcons}), it is 
clear that tensors in the left hand side of equations in (\ref{tqcons}) do not correspond to 
free energy-momentum tensors, and along with $Q^\mu$ they are obtained somehow arbitrarily by 
division of (\ref{ttotcons}). In fact, equations in (\ref{tqcons}) are inspired by 
perturbation theory in which for each perturbative order, the right hand sides of these 
equations are estimated by using quantities from one perturbative order lower. Thus, they 
constitute an iterative set of equations from zero order (free) model in which $Q^\mu = 0$, 
up to higher orders. This approach is not suitable for dark energy where we ignore, not only 
interactions but also the free model. Therefore, a more general expression should be used: 
\be
T^{\mu\nu}_{m~;\nu} = -Q_m^\mu, \quad T^{\mu\nu}_{\varphi~;\nu} = -Q_\varphi^\mu, \quad 
T^{\mu\nu}_{int~;\nu} = Q_m^\mu + Q_\varphi^\mu \label{tqmcons}
\ee
In these equations matter and dark energy tensors $T^{\mu\nu}_m$ and $T^{\mu\nu}_\varphi$ have the 
same expression as in the absence of interaction, but with respect to fields which are not 
free. These expressions can be justified by considering the Lagrangian of the model. In 
Einstein frame the Lagrangian for a weakly interacting system can be divided to free and 
interaction parts:
\be
{\mathcal L} = \sum_i {\mathcal L}_i + {\mathcal L}_{int} \label{ltot}
\ee
Considering only local interactions, in the dynamics equations for the fields partial 
derivative of ${\mathcal L}_{int}$ with respect to each field determines the interaction term. 
Dynamic equations can be related to energy-momentum conservation equations 
(\ref{tqcons})~\cite{houriquin}. Therefore, interaction currents $Q_m^\mu$ and $Q_\varphi^\mu$ 
are generated by partial derivatives of ${\mathcal L}_{int}$ with respect to the corresponding 
field.

In the previous section we explained that the scalar field in scalar-tensor modified gravity 
models is related to a dilaton. Consequently, the interaction term is proportional to the 
trace of matter, see equation (\ref{grgmener}) for an explicit example of 
$\mathrm {f}(R)$ models. In this case there is no interaction between scalar field and 
relativistic particles, and it can be shown that $Q_m^\mu = -Q_\varphi^\mu$~\cite{frbean}, i.e. 
$T^{\mu\nu}_{int~;\nu} = 0$ and conservation equations in (\ref{tqcons}) can be used. Interaction 
current $Q^\mu$ for these models can be written as:
\be
Q^\mu = {\mathcal C}(\varphi) T_m \partial^\mu \varphi \label{mgcurrent}
\ee
where $T_m = g_{\mu\nu} T^{\mu\nu}_m$. In $f(R)$ models the coupling ${\mathcal C}$ is a constant. 
Here we consider $\varphi$-dependence to cover more general cases. Some authors have also 
considered $Q^\mu \propto T_m u^\mu_m$ for interacting quintessence models~\cite{quinintterm}. 
In fact, the interaction current of interacting dark energy models in literature is usually 
considered to be $Q^0 \propto \rho_m = T_m$ for cold dark matter i.e. similar to what is 
obtained for $f(R)$ modified gravity models~\cite{frbean}. However, giving the fact that these 
models share some important properties with modified gravity models, such as the absence of 
interaction between relativistic matter and scalar field, we classify them in the modified 
gravity category. In fact, interactions in interacting quintessence models can be more diverse 
than this simple case. In the rest of this section we describe how they can be formulated 
without considering their details. 

In the context of quantum field theory, the Lagrangian ${\mathcal L}$ can be easily written 
for various types of fields and their interactions, see e.g.~\cite{houricondens}. But these  
formulations are usually complicated, and are necessary if the microphysics of dark energy 
models is studied. There are various ways to write ${\mathcal L}$ and/or $T^{\mu\nu}$ with 
respect to macroscopic quantities which are apriori measurable from cosmological observations. 
For instance, one can use a fluid description for components. The Lagrangian of a fluid is 
defined as~\cite{fluidlagrang}:
\bea
&& {\mathcal L}_f = \frac{1}{2} (P+\rho) g_{\mu\nu} u^\mu u^\nu + \frac{1}{4} (P-\rho) 
g_{\mu\nu}g^{\mu\nu} + \frac{1}{2} g_{\mu\nu} \Pi^{\mu\nu} \label{lfluid} \\
&& \rho \equiv K + V, \quad P \equiv K - V \label{rhopdef}
\eea
where $K$ and $V$ are respectively kinetic and potential energy, and $\Pi^{\mu\nu}$ is the 
traceless shear tensor. Note that if we impose the traceless condition on the Lagrangian, the 
last term in the right hand side of equation (\ref{lfluid}) becomes zero. Therefore, this term 
must be considered as a Lagrange multiplier, and traceless condition is imposed after 
determination of $T^{\mu\nu}_f$~\cite{fluidlagrang}. It is easy to check that the Lagrangian 
${\mathcal L}_f$ leads to the familiar expression for the energy-momentum tensor of a fluid:
\be
T^{\mu\nu} \equiv \frac{2}{\sqrt{-g}}\biggl [\frac{\partial (\sqrt{-g} {\mathcal L})}
{\partial g_{\mu\nu}} - \partial_\rho \biggl (\frac{\partial (\sqrt{-g} {\mathcal L})}
{\partial (\partial_\rho g_{\mu\nu})} \biggr )\biggr ], \quad\quad
T^{\mu\nu}_f = (\rho+P) u^\mu u^\nu - g^{\mu\nu}P + \Pi^{\mu\nu} \label{tfluid}
\ee
Transformation of a Lagrangian written with respect to fields to a fluid description is easy, 
and one can determine the energy-momentum of interaction $T^{\mu\nu}_{int}$ and the current $Q^\mu$ 
defined in (\ref{tqmcons}) directly and without any ambiguity, see Appendix \ref{app:c}. 
However, their descriptions as a function of density and pressure of the fluid depend on the 
self-interaction potential $V(\varphi)$. For instance, a Higgs-like interaction between a scalar 
and a fermion $\propto \varphi \bar\psi \psi$ is described as $\propto (\rho_\psi - P_\psi)
(\rho_\varphi - P_\varphi)^{1/2}$ if $V(\varphi) \propto \varphi^2$, and as $\propto 
(\rho_\psi - P_\psi) (\rho_\varphi - P_\varphi)^{1/4}$ if $V(\varphi) \propto \varphi^4$. Therefore, 
when the objective is a general parametrization of interactions without considering details 
of the underlying model, this type of description is not very suitable. 

A more serious problem of fluid description of interaction Lagrangian is the fact that 
conservation equations in (\ref{tqmcons}) are equivalent to field equations and can be 
obtained from them~\cite{houriquin}. Therefore, they do not contain quantum processes such 
as decay and scattering. It is well known that the Boltzmann equation plays the 
role of intermediate between quantum and classical description of interacting 
systems~\cite{boltzquantum,boltzquantum1,boltzquantum2,boltzquantum3,boltzquantum4}. In this 
case, components are defined by their phase space distribution $f(p,x)$ where $p$ and $x$ are 
respectively momentum and spacetime coordinates. Interactions are included as collision terms 
in the right hand side of the Boltzmann equation~\cite{boltzcoll0,boltzcoll1,boltzcoll2}, from 
which one can obtain energy-momentum and number conservation equations directly:
\bea
&& p^\mu \partial_\mu f_i (p,x) - \Gamma^\mu_{\nu\rho} p^\nu p^\rho \frac{\partial f_i}
{\partial p^\mu} \equiv L[f_i] = C_i(p,x) \label{boltzeq} \\
&& n^\mu_{;\mu} = \int d\bar{p}~C_i(p,x), \quad d\bar{p} \equiv \frac{{\mathbf g}}{(2\pi)^3} 
d^4p \delta (E^2 - \vec{p}^2 - m_i^2) \label{boltzn} \\
&& T^{\mu\nu}_{i~;\nu} = \int d\bar{p}~p^\mu C_i(p,x) \label{boltzt}
\eea
where ${\mathbf g}$ is the number of internal degrees of freedom (e.g. spin) of species $i$. 
Conservation equations (\ref{boltzn}) and (\ref{boltzt}) are obtained by using the following 
property of the Boltzmann operator $L$ defined in (\ref{boltzeq}), see e.g.~\cite{boltzcoll1}:
\be
\biggl[\int d\bar{p}~p^{\mu}p^{\mu_1}p^{\mu_2}\ldots p^{\mu_n} f(p,x)\biggr ]_{;\mu} = 
\biggl[\int d\bar{p}~p^{\mu_1}p^{\mu_2}\ldots p^{\mu_n} L[f(p,x)]\biggr ] \label{boltzint}
\ee
Collisional terms can be written by using cross-sections of interactions which can be determined 
separately from the quantum formulation of the model~\cite{boltzcoll0,houridecay}. In the 
context of interacting dark energy models, the simplest examples of collisional terms are 
elastic scattering between dark matter and dark energy and slowly decay of dark matter with a 
small branching ratio to dark energy~\cite{houriquin}\footnote{In models where energy is 
transferred from dark energy to dark matter, the interaction must be nonlinear and very 
sophisticated such that a very light quintessence field be able to produce massive dark matter 
particles. At present no fundamental description for such models is available.}. Note that 
we assume no interaction between dark energy and visible matter and radiation. For these 
interactions the collisional terms are:
\bea
C_m (p,x) &=& -\Gamma_m m_m f_m(p,x) - f_m(p,x) \int d\bar{p}_\varphi~f_\varphi (p_\varphi,x)~
A_k(p,p_\varphi)~\sigma_{m\varphi} (p,p_\varphi) + \nonumber \\
&& \int d\bar{p}_m~d\bar{p}_\varphi~f_m(p_m,x)~f_\varphi (p_\varphi,x)~A_k (p_m,p_\varphi)
\frac{d\sigma_{m\varphi}(p_m,p_\varphi,p)}{d\bar{p}} \label{cmatter}\\
C_\varphi (p,x) &=& \Gamma_m m_m \int d\bar{p}_m~f_m (p_m,x)~\frac{d{{\mathcal M}(p_m,p)}}
{d\bar{p}} - f_\varphi (p,x) \int d\bar{p}_m~f_m (p_m,x)~A_k(p,p_m) \nonumber \\
&& ~\sigma_{m\varphi} (p_m,p) + \int d\bar{p}_m~d\bar{p}_\varphi~f_m(p_m,x)~
f_\varphi (p_\varphi,x)~A_k (p_m,p_\varphi)~\frac{d\sigma_{m\varphi} (p_m,p_\varphi,p)}
{d\bar{p}} \label{cde} \\
A_k (p_1,p_2) &\equiv& [(p_1.p_2)^2 - m_1^2 m_2^2]^{\frac{1}{2}} \label{akin}
\eea
where $\Gamma_m$ is the total decay width of dark matter, ${\mathcal M} (p_m,p)$ is the 
multiplicity of $\varphi$ with momentum $p$ in the decay remnants of dark matter particles with 
momentum $p_m$, and $\sigma_{m\varphi} (p_m,p_\varphi)$ is the total cross-section of interaction 
between dark matter and dark energy with momentum $p_m$ and $p_\varphi$, 
respectively\footnote{Note that although dark energy is a condensate i.e. its {\it particles} 
have the same energy, presumably zero momentum, a general condensate state can contain very large 
number of {\it particles} in different energy levels, see~\cite{houricondens} for more details.}.

The disadvantage of this approach is that it needs phase space distribution of components which 
is not always available, specially for dark energy. Moreover, the absence of an explicit 
description for the Lagrangian means that the total energy-momentum tensor needed for 
determining Einstein equations and metric evolution, can be obtained only by solving equation 
(\ref{boltzt}) for all components. These equations are differentio-integral and usually don't 
have analytical solution. Thus, in practice interacting models can be studied 
only numerically, otherwise one needs to consider some approximation. For instance, dark 
energy interaction with matter must be very weak. Thus, $|T^{\mu\nu}_{int}| \ll |\sum_i 
T^{\mu\nu}_i|$. Therefore, we can neglect its contribution in the total energy-momentum tensor 
and Einstein equations\footnote{In some dark energy models such as {\it early dark energy} it 
is assumed that the density of dark energy at high redshifts is much larger, and only at low 
redshifts it is reduced. Although at redshifts relevant for dark energy surveys cosmology 
must be very close to $\Lambda$CDM, one must be aware that in many models of this type, the 
approximation of weak interaction can be applied only at low redshifts. It is also expected 
that these models leave a detectable signature on the CMB spectrum~\cite{earlyde,earlyde1}.}. 
As for the integration of collision term in equations 
(\ref{boltzn}) and (\ref{boltzt}), under some physically motivated assumptions they can be 
simplified and integrated. For instance, when dark matter is assumed to be 
an scalar, the expression for the scattering cross-section is very simple, see 
e.g.~\cite{houriquin}. It is simply proportional to the coupling constant and delta functions 
for energy-momentum conservation. It is expected that the mass of quintessence field be very 
small, specially much smaller than the mass of dark matter. The momentums of both components 
are also expected to be small. In this case their distribution at large momentums is strongly 
suppressed, and the cross-section around the peak of distribution can be considered to be 
approximately constant. Under these simplifications, it is easy to see that scattering term 
in the right hand side of (\ref{boltzt}) is proportional to integrals of the form:
\bea
&& \int d\bar{p}_1~d\bar{p}_2~P_1^\mu~f_1(p_1,x)~f_2(p_2,x) = n_1^\mu \int d\bar{p}_2~f_2(p_2,x) 
\approx \frac{n_1^\mu~u_\rho}{m_2}\int d\bar{p}_2~p_2^\rho ~f_2(p_2,x) = 
\frac{n_1^\mu~u_{2\rho}~n_2^\rho}{m_2} \nonumber \\
&& \label{ffintapprox} \\
&& u_i^\mu \equiv \frac{n_i^\mu}{|n_i|}, \quad n_i^\mu \approx \frac{u_{i\nu} T_i^{\mu\nu}}
{m_i} = \frac{\rho u_i^\mu}{m_i} \label{urelate}
\eea
where $n_i^\mu$ and $ u_i^\mu$ are number density and velocity of species $i$, respectively. 
Approximate expression for $n^\mu$ in (\ref{urelate}) is valid when the distribution in momentum 
space is concentrated around a peak. Using similar approximations the decay terms in the right 
hand side of (\ref{boltzt}) can be also described as a function of velocity and number vectors. 
Finally, after grouping all the constant or approximately constant factors together,  
energy-momentum conservation equations for dark matter and dark energy can be written as:
\bea
&& T^{\mu\nu}_{m~;\nu} \approx -L_m n_m^\mu + A_{ms} n_m^\mu u_{\varphi\rho} n_\varphi^\rho 
\equiv Q_m^\mu \label{tmconsmapprox} \\
&& T^{\mu\nu}_{\varphi~;\nu} \approx L_\varphi n_m^\mu + A_{\varphi s} n_\varphi^\mu 
u_{m\rho} n_m^\rho \equiv Q_\varphi^\mu \label{tdeconsmapprox}
\eea
where constants $L_i$ and $~A_{is}$ are decay width and scattering amplitude for species $i$.
In the rest of this work we use these equations as an approximation for energy-momentum 
conservation equations irrespective of dark matter type (spin) and details of interaction 
between two dark components. They affect constants $L_i$ and $~A_{is}$ which are used as 
parameters. One can also add dark matter self-annihilation term to (\ref{tmconsmapprox}). But, 
it is easy to show that self-annihilation is proportional to $|n_m|^2$. Thus, it is 
significant only in dense regions i.e. at small spatial scales such as the central region of 
dark matter halos, which are in nonlinear regime and are not studied in the present work. 
Here we only consider homogeneous and linear perturbations. Therefore, the effect of 
annihilation is negligible. We remind that equation (\ref{tmconsmapprox}) is not restricted 
to cold dark matter and can be also used for relativistic matter, for instance neutrinos in 
early universe, or a hot component at low redshifts. 

Although in the rest of this work we consider the interaction terms described in this 
section, for what concerns the study of differences between modified gravity and interacting 
quintessence models, the formulation of anisotropies and discrimination methods explained 
in the next two sections can be applied to other choices of interactions. It is enough to find 
an interaction current similar to what we have found for decay and scattering above and add 
them to the right hand side of equations (\ref{tmconsmapprox}) and (\ref{tdeconsmapprox}).

\section{Cosmology and evolution of anisotropies} \label{sec:perturb}
In this section we first determine ${\mathcal F}_i$ coefficients defined in 
Sec. \ref{sec:frieman} for both modified gravity and quintessence models according to 
interaction currents and energy-momentum conservation equations obtained in the previous 
section. Then, we consider the effect of interactions on the evolution of anisotropies, and 
describe how interaction parameters can be extracted from data.

\subsection{Interaction coefficients in the Friedman equation} \label{sec:fi}
\subsubsection {Modified gravity} \label{sec:fimg}
Using the energy momentum conservation equation (\ref{tqmcons}) and the interaction current for 
modified gravity models, the scalar field equation and the evolution equation of the 
homogeneous matter density can be determined as the followings~\cite{frbean}:
\bea
&& \bar{\varphi}'' + 2{\mathcal H} \bar{\varphi}' + a^2 V_\varphi (\bar{\varphi}) = 
a^2 {\mathcal C} (\bar{\varphi}) \sum_i(\bar{\rho}_i -3 \bar{P}_i), \quad {\mathcal H} = 
\frac{a'}{a} \label {modgrphi}\\
&& \bar{\rho}'_i + 3{\mathcal H} (\bar{\rho}_i + \bar{P}_i) = {\mathcal C} (\bar{\varphi}) 
\bar{\varphi}' (\bar{\rho}_i - 3\bar{P}_i) \quad i=m,~b,~h\label {modgrrho}
\eea
where barred quantities are homogeneous components, $\varphi$ in subscript means derivative 
with respect to $\varphi$. Note that here we have generalized the original calculation 
in~\cite{frbean} by considering a $\varphi$-dependent ${\mathcal C} (\bar{\varphi})$ 
coefficient in the right hand side of these equations to cover a larger class of modified 
gravity models, see e.g.~\cite{deint}. For $\mathrm {f}(R)$ models ${\mathcal C} = 
\sqrt {4\pi G/3}$~\cite{frbean}. Equations (\ref{modgrphi}) and (\ref{modgrrho}) are coupled 
and an analytical solution can not be found without considering an explicitly $V(\varphi)$. 
Therefore, to solve the equation for $\bar{\rho}$, which is in fact the only directly 
observable quantity, we simply consider the right hand side of the equation as a time-dependent 
source. The solution of equation (\ref{modgrrho}) can be written as:
\be
\bar{\rho}_i (z) = \bar{\rho}_i (z_0) (1+z)^{3(1+w_i)} e^{(1-3w_i) F(\bar{\varphi})}, \quad 
F(\varphi) \equiv \int {\mathcal C} (\bar{\varphi}) d\varphi, \quad i = m,b,h \label {mgrhohomo}
\ee 
where $w_i \equiv \bar{P}_i/\bar{\rho}_i$ for all species except dark energy are assumed to be 
constant and are given in equation (\ref{fmgdef}). Comparing this solution with 
(\ref{friedmanintde}) we find:
\be
{\mathcal F}_i(z) = e^{(1-3w_i) F(\bar{\varphi} (z))} \approx 1 + (1-3w_i) F(\bar{\varphi} (z)) 
\label{mgfi}
\ee
In $\mathrm {f}(R)$ models ${\mathcal C} (\bar{\varphi}) = \sqrt {4\pi G/3} 
\equiv C$~\cite{frbean} is a constant, thus:
\be 
F(\bar{\varphi}) = C\bar{\varphi}(z). \label{fphibar}
\ee
Using transformation from Jourdan frame to Einstein frame $\bar{\varphi}(z) = 
\ln (\mathrm {f}_R(z) + 1) / 2C$~\cite{frbean}, one can relate ${\mathcal F}_i(z)$ to 
$\mathrm {f}_R$:
\be
{\mathcal F}_i(z) \approx 1 + \frac{(1-3w_i)}{2} \ln (\mathrm {f}_R + 1) \approx 
(1 + \mathrm{f}_R)^{-\frac{(1-3w_i)}{2}} \label{fiphibar}
\ee
The approximate expression in (\ref{fiphibar}) is the same as equation (\ref{fmgdef}). Note 
that in (\ref{mgrhohomo}) all constant coefficients including 
$(1 + \mathrm{f}_R(z_0))^{-\frac{(1-3w_i)}{2}}$ are included in $\bar{\rho}_i (z_0)$. Apriori one 
can test the presence of a $\mathrm {f}(R)$ modified gravity by measuring simultaneously 
${\mathcal F}_m(z)$, ${\mathcal F}_h(z)$, and equation of state of dark energy from equation 
(\ref{mgdedens}). In fact in this equation if we neglect the last term that depends on the time 
derivative, the effective dark energy density becomes:
\be
\rho_{de} \approx \frac{3}{8\pi G}~\frac{\mathrm {f}_R}{1+\mathrm {f}_R}(-\frac{d\ln 
\mathrm {f}(R)}{6~dR} - \frac {R}{6}) \label{mgrdeapp}
\ee
To be consistent with observations $\mathrm {f}(R)$ cannot be a fast varying function of $R$. 
Therefore, the dominant term in (\ref{mgrdeapp}) is the term proportional to $R$ which makes the 
relation between $\rho_{de}$, $R$, and $\mathrm {f}(R)$ very simple. Other ${\mathcal F}_i$'s and 
evolution of corresponding densities have also known expressions, notably 
${\mathcal F}_h(z) = 1$. Therefore, apriori simultaneous fitting of these quantities can test 
$\mathrm {f}(R)$ modified gravity models. More generally, in modified gravity models dark 
energy term in the Friedman equation is an effective contribution generated from 
non-conventional interaction between matter and gravity. Therefore, it is more correlated to 
matter than in $\Lambda$CDM or (interacting)-quintessence models. In the former apriori there 
is no correlation in the dark sector, and in the latter case the interaction can be very small 
and is only necessary for reducing fine-tunings and making the model more natural. 
Similar correlation tests can be performed for other modified gravity models too. Evidently, 
giving the small deviation of dark energy from a cosmological constant, the measurements and 
calculation of correlations are not trivial tasks. Furthermore, the discrimination must be 
cross-checked by using anisotropies for distinguishing between dark energy models, explained 
in Sec. \ref{sec:perturbint}.

\subsubsection {Interacting quintessence} \label{sec:fiquin}
In the same way, we can determine ${\mathcal F}_i$ coefficients for (interacting)-quintessence 
using equation (\ref{tmconsmapprox}). We replace $n^\mu$ with approximation (\ref{urelate}) 
and include $1/m$ factors in the $L$ and $A_S$ coefficients. After these simplifications, the 
evolution equation for the density of interacting quintessence models becomes:
\be
\bar{\rho}'_i + 3{\mathcal H} (\bar{\rho}_i + \bar{P}_i) = -L_i a \bar{\rho}_i + A_{si} a 
\bar{\rho}_i \bar{\rho}_\varphi \label{quinrhohomoeq}
\ee
where $i$ indicates any cold matter or relativistic species that interact with quintessence 
field\footnote{If species $i$ has interaction with another component, for instance is scattered 
by another species, we can add a second scattering term to (\ref{quinrhohomo}). The best example 
is the scattering of photons or neutrinos by baryons. Here for the sake of simplicity we neglect 
such interactions which are not the main concern of this work. However, in a full formulation of 
the problem they should be considered, specially if they can mimic an interaction with dark 
energy.}. A clear difference between interaction term in (\ref{quinrhohomoeq}) and 
(\ref{modgrrho}) is that the former does not explicitly depend on the scalar field, and 
therefore we do not need to know and solve a field equation similar to (\ref{modgrphi})
\footnote{For $f(R)$ modified gravity in which ${\mathcal C}$ is constant $\bar{\varphi}'$ in 
(\ref{modgrrho}) can be replaced by an expression depending on density and pressure, and there 
is no need for solving field equation of the scalar field either.}. The solution of this 
equation and corresponding ${\mathcal F}_i$'s are:
\bea
\bar{\rho}_i (z) &=& \bar{\rho}_i (z_0) (1+z)^{3(1+w_i)} \exp \biggl (L_i (\tau (z) - 
\tau (z0)) + A_{si}\int dz \frac{\bar{\rho}_\varphi (z)}{(1+z) H (z)} \biggr ) 
\label{quinrhohomo} \\
{\mathcal F}_i (z) &=& \exp \biggl (-L_i (\tau (z) - \tau (z_0)) + A_{si}\int dz 
\frac{\bar{\rho}_\varphi (z)}{(1+z) H (z)} \biggr ) \approx 1 + L_i (\tau (z_0) - \tau (z)) + 
\nonumber \\
&& A_{si}\int_{z_0}^z dz \frac{\bar{\rho}_\varphi (z)}{(1+z) H (z)} \label{quinfi}
\eea
where $\tau (z)$ is the age of the Universe at redshift $z$. Note that even in absence of 
expansion, the density of dark matter at high redshifts is higher if $L_i > 0$.

Along with consistency relation explained above for modified gravity models, explicit dependence 
of (\ref{quinfi}) on measurable quantities $\bar{\rho}_\varphi (z)$ and $H (z)$ apriori allows to 
discriminate between interacting quintessence and modified gravity models. Note that the prior 
knowledge about the evolution of these quantities are mandatory for distinguishing the 
underlying model and without such information one cannot single out any of these models.

\subsection{Matter perturbations in interacting dark energy cosmologies} 
\label{sec:perturbint}
Although dark energy influences the evolution of perturbations mainly through quantities 
related to the homogeneous component - background cosmology - the study of anisotropies can be 
a powerful mean both for measuring the equation of state and for discriminating between 
candidate models. Standard candles, such as supernovae type Ia, allow direct measurements of 
distances, and thereby cosmological parameters. However, they are rare events, can deviate from 
being standard due to absorption or late detection~\cite{anabsorbe}, sub-types, and dependence 
of their light curve on other properties such as metallicity, mass, and magnetic field of 
progenitors~\cite{snphysics}. Determination of dark energy properties from evolution of 
perturbations provides additional information and a mean for cross-check of the two methods.

Matter perturbations in presence of an interacting dark energy~\cite{quinpertub} and in 
$\mathrm {f}(R)$ modified gravity models~\cite{frgr,frgr1,frbean,frsilvestri} have been 
calculated by various authors, thus here we do not repeat them and simply use their results. 
Our main objective is to find and discuss features that can be used for discrimination between 
dark energy models.

Considering only scalar perturbations, we define the first-order metric in conformal gauge as 
the following:
\be
ds^2 = a^2 (\eta) [(1+2\psi (\mathbf{x})) d\eta^2 - (1-2\phi(\mathbf{x})) \delta_{ij} dx^i dx^j] 
\label{metric}
\ee
As we mentioned in the Introduction, for modified gravity models we write evolution equations 
in Einstein frame. Thus, here only their interaction terms distinguish them from quintessence 
models. 

We use fluid description for both matter and dark energy. After linearizing energy-momentum 
conservation equations and taking their Fourier transform with respect to spatial coordinates, 
evolution equations for density and velocity perturbations of matter component $i$ and dark 
energy can be written as:
\bea
\delta \rho'_{(i)} + 3 {\mathcal H}\delta \rho_{(i)} (1 + C_{s(i)}^2) + (1+w_{(i)}) 
\bar{\rho}_{(i)} (3 \phi'- ik_j v_{(i)}^j) = \delta Q_{(i)0}&& \label{deltarhoeq} \\
((1+w_{(i)}) \bar{\rho}_{(i)} v_{(i)j})' + 4 {\mathcal H} (1+w_{(i)}) \bar{\rho}_{(i)} v_{(i)j} - 
i k_{(i)} C_{s(i)}^2\delta\rho_{(i)} - i k_l \Pi^l_{(i)j} - ik_j (1+w_{(i)}) \bar{\rho}_{(i)} \psi = 
\delta Q_{(i)j} && \label{deltaveq} \\
\delta \rho'_\varphi + 3 {\mathcal H}\delta \rho_\varphi (1 + C_{s\varphi}^2) - 
(1+w_\varphi) \bar{\rho}_\varphi (3 \phi'- ik_j v_\varphi^j) = \delta Q_{\varphi 0} &&
\label{deltarhophieq} \\
((1+w_{(i)}) \bar{\rho}_\varphi v_{\varphi j})' + 4 {\mathcal H} (1+w_\varphi) 
\bar{\rho}_\varphi v_{\varphi j} - i k_j C_{s\varphi}^2 \delta \rho_\varphi - 
i k_l \Pi^l_{\varphi j} - ik_j (1+w_\varphi) \bar{\rho}_\varphi \psi = \delta Q_{\varphi j} &&
\label{deltavphieq}
\eea
where $C_{s(i)}^2 \equiv \delta P_{(i)}/ \delta\rho_{(i)}$ is the speed of sound for species $i$
\footnote{To prevent confusion between spacetime indices and indices indicating the species, 
when there is a risk of confusion we put the latter inside brackets}, $v_{(i)}$ is its velocity, 
and $\Pi^l_{(i)j}$ is its anisotropic shear. The perturbation of interaction current for modified 
gravity and quintessence models derived from (\ref{mgcurrent}), (\ref{tmconsmapprox}) and 
(\ref{tdeconsmapprox}) are as the followings:

{\bf Modified gravity:}
\bea
\delta Q_{(i)0} &=& \rho_{(i)} \biggl [(1-3w_{(i)}) {\mathcal C}_\varphi (\bar{\varphi}) 
\bar{\varphi}' \delta \varphi + {\mathcal C} (\bar{\varphi}) \biggl ((1-3w_{(i)}) \delta 
\varphi' + (1 -3 C_{s(i)}^2) \bar{\varphi}' \delta_{(i)} \biggr) \biggr ] \nonumber \\
&=& -\delta Q_{\varphi 0} \label{mgqzero} \\ 
\delta Q_{(i)j} &=& ik_j {\mathcal C}(\bar{\varphi}) \bar {\rho}_{(i)}
(1 - 3w_{(i)}) \delta \varphi = -\delta Q_{\varphi j}, \quad i = m,~b,~h \label{mgqi}
\eea
{\bf Interacting quintessence:}
\bea
\delta Q_{(i)0} &=& -a L_{(i)} (\delta \rho_{(i)} + \bar{\rho}_{(i)} \psi) + a A_{s(i)} 
\biggl [\bar{\rho}_\varphi \delta \rho_{(i)} + \bar{\rho}_{(i)} (\delta \rho_\varphi + 
\bar{\rho}_\varphi \psi)\biggr ] \label{quinqzeromatter} \\
\delta Q_{(i)j} &=&  av_{(i)j} (-L_{(i)} \bar{\rho}_{(i)} + A_{s(i)} \bar{\rho}_{(i)} 
\bar{\rho}_\varphi) \label{quinqimatter} \\
\delta Q_{\varphi 0} &=& aL_\varphi (\delta \rho_{(i)} + \psi \bar{\rho_{(i)}}) + a A_{\varphi s} 
\biggl [\delta\rho_\varphi \bar{\rho}_{(i)} + \bar{\rho}_\varphi \delta\rho_{(i)} - 
\bar{\rho}_{(i)}\delta\rho_\varphi \biggl (\frac{1 + C_{s\varphi}^2}{1 + w_\varphi} + \psi \biggr ) 
\biggr ] \label{quinqzerode} \\
\delta Q_{\varphi j} &=&  av_{(i)j} (L_\varphi \bar{\rho}_{(i)} + A_{s(i)} \bar{\rho}_{(i)} 
\bar{\rho}_\varphi), \quad i = \text{All matter interacting with $\varphi$} \label{quinqide}
\eea
In (\ref{mgqzero}), ${\mathcal C}_\varphi$ is the derivative of ${\mathcal C}(\varphi)$ with 
respect to $\varphi$. To obtain these equations we have used the following definition and 
properties:
\bea
&& \rho_\varphi \equiv u_{(\varphi) \mu} u_{(\varphi) \nu}T_\varphi^{\mu\nu} = 
\frac{1}{2} \partial_\mu \varphi \partial^\mu \varphi + V (\varphi) \label{rhophidef} \\
&& P_\varphi \equiv \frac{1}{2} \partial_\mu \varphi \partial^\mu \varphi - V (\varphi) 
\label {pphidef} \\
&& u_(\varphi)^\mu \equiv \frac{\partial^\mu \varphi}{\partial_\nu \varphi \partial^\nu 
\varphi} = \frac{\partial^\mu \varphi}{(\rho_\varphi + P_\varphi)^{\frac{1}{2}}} 
\label{uphidef} \\
&& \frac{\delta \rho_\varphi + \delta P_\varphi}{\rho_\varphi + P_\varphi} = 
2 \biggl [\frac{a \partial^0(\delta \varphi)}{(\bar{\rho}_\varphi + \bar{P}_\varphi)^
{\frac{1}{2}}} + \psi \biggr ] \label{deltaphip}
\eea
Evidently, these equations are valid for both modified gravity and interacting quintessence. 
They are also highly coupled, thus it is impossible or very difficult to find an analytical 
solution for them. To complete evolution equations for modified gravity, we also need the 
evolution of $\delta \varphi$. This can be obtained by expanding the field 
$\varphi = \bar{\varphi} + \delta \varphi$ and using the covariant field equation, see 
e.g. ~\cite{houricondens}:
\bea
&& \frac{1}{\sqrt{-g}} \partial_\mu (\sqrt{-g} g^{\mu\nu} \partial_\nu \varphi) + 
V_\varphi (\varphi) = {\mathcal C} (\varphi) T_m \label{covarfieldeq}\\
&& \delta \varphi'' + 2 {\mathcal H}\delta \varphi' + \psi'(\bar{\varphi}' +  2 {\mathcal H}
\bar {\varphi}) + \psi (\bar{\varphi}'' +  2 {\mathcal H} \bar {\varphi}' + 2 \frac{a''}{a} 
\bar {\varphi}) - (k^2 - \frac{a''}{a} + V_{\varphi\varphi}) \delta\varphi = 
{\mathcal C} (\varphi) \delta T_m + C_\varphi \bar {T} \delta\varphi \nonumber \\
&& \label{deltaphieq}
\eea
As we mentioned in previous sections, solving these equations is not the main aim of present 
work. Our goal is to single out differences of these models that can be used for discriminating 
them from other models. For instance, in modified gravity models the perturbation of 
interaction current does not 
depend on the metric perturbations $\psi$ and $\phi$. By contrast, in interacting dark energy 
the current perturbation depends on the metric perturbation and it is easy to that:
\be
\frac{\text {term} \propto \psi}{\text {term} \propto \delta} = 1 \label{quinpsidelta}
\ee
Because according to observations $\delta \varphi$, $\varphi'$, and $\delta \varphi'$ are 
very small, in both models the terms proportional to $\delta_i \equiv \delta \rho_i/\rho_i$ 
are dominant. In this case, it is easy to see that for modified gravity $\delta Q_{(i)0} \propto 
{\mathcal C} (\varphi)$ and for interacting quintessence $\delta Q_{(i)0} \propto (-L_\varphi + 
A_{s(i)} \bar{\rho}_\varphi)$. Although apriori these quantities evolve differently, both of 
them are expected to vary very slowly. Thus, it is not possible to distinguish them, specially 
in a model independent way. Other properties such as (\ref{quinpsidelta}) cannot be used 
directly either. Nonetheless, they influence the growth rate $\propto \delta'_i/\delta_i$, 
density power spectrum, and density-velocity correlations, etc.. In the next section we discuss 
how these measurable quantities can be related to interaction current, and thereby allow to 
discriminate between modified gravity and quintessence models.

Perturbation equations (\ref{quinqzeromatter}) to (\ref{quinqide}) depend on metric 
perturbations $\psi$ and $\phi$, and their time derivatives. These quantities can be determined 
from Einstein equations for perturbations (see e.g.~\cite{perturb}):
\bea
&& k^2 \phi + 3 {\mathcal H} (\phi' + {\mathcal H} \psi) = 4\pi G a^2 \sum_i \delta \rho_i 
\label{einst00} \\
&& k^2 (\phi' + {\mathcal H} \psi) = -4\pi G a^2 \sum_i i k_j v^j_{(i)} (\bar{\rho}_{(i)} + 
\bar{P}_{(i)}) \label{einst0i} \\
&& \phi'' + {\mathcal H} (\psi' + 2\phi') + (~\frac{2a''}{a} - \frac{{a'}^2}{a^2}) \psi + 
\frac{k^2}{3} = -4\pi G a^2 \sum_i \delta P_i \label{einstii} \\
&& k^2 (\phi - \psi) = -12 \pi G a^2 \sum_i (k_j k^l - \frac{1}{3} \delta^l_j) \Pi^j_{(i)l} 
\label{einstij}
\eea
Note that in these equations the interaction energy is neglected. The reason is that we need 
$T_{int}^{\mu\nu}$, which in the phenomenological description of interactions is not known. 
Nonetheless, its omission in equations (\ref{einst00}) to (\ref{einstij}) should not induce 
large errors because present observations show that any non-gravitational interaction between 
various constituents of the Universe - if any - must be very small, and therefore this 
approximation is justified. Metric perturbations $\psi$ and $\phi$ cannot be directly observed, 
except through lensing. Otherwise, they can be extracted from these equations when 
density-density and density-velocity correlations, and induced anisotropic shear $\Pi^j_{(i)l}$ 
are determined from LSS data.

Although phenomenological interaction currents (\ref {tmconsmapprox}) and 
(\ref {tdeconsmapprox}) are inspired from well understood scattering of particles, one cannot 
rule out other types of interaction. Even for these cases apriori one should be able to write 
equations similar to (\ref {quinqzeromatter})-(\ref {quinqide}), and (\ref {quinpsidelta}). 
The fact that the latter relations are independent of the strength of the coupling between dark 
energy and matter proves that finding a different proportionality between $\psi$ and $\delta$ 
terms would be a clear signature of an unusual quintessence model, e.g. one with a non-minimal 
interaction with gravity. Evidently, such measurements are not easy. Nonetheless, with the huge 
amount of data expected from near future surveys and their better precision, more accurate 
measurements of parameters should be possible, and precision analysis necessary for detailed 
examination of dark energy models should be achievable.

\section{Estimation of forecast precision for surveys} \label{sec:forecast}
In this section we first describe how in practice the background cosmology parameters defined 
in Sec. \ref{sec:frieman} are calculated. Their uncertainties determine how well a survey can 
discriminate between modified gravity and (interacting)-quintessence models, independent of the 
data type or observation method. Then, we calculate and parametrize the evolution equation of 
the growth rate of matter anisotropies and discuss its measurement uncertainty. As an example 
we make an order of magnitude estimate for the expected uncertainty of these quantities for 
the Euclid mission~\cite{euclid}. As we mentioned in the Introduction, a proper forecast needs 
detailed study of observational effects and uncertainties which is out of the scope of present 
work.

\subsection{Discriminating between a cosmological constant and other models}
\label{sec:forcc}
As we discussed in Sec. \ref{sec:errordiscrim}, discrimination ability of surveys between a 
cosmological constant and a redshift dependent dark energy can be evaluated by using the 
function $\Theta (z)$ defined in (\ref{deftheta}). To calculate the quantity $\Theta$ and its 
uncertainty, we need to know uncertainties of the estimation of effective background cosmological 
parameters. The function $\Theta$ depends on $\Omega_{eff}^{(H)}$, $w_{eff}^{(H)}(z)$, 
$\Omega_{eff}^{(A)}$, and $w_{eff}^{(A)}(z)$, effective dark energy fractional density and equation 
of state dark energy determined, by fitting $H(z)$ and $A(z)$, respectively. By measuring 
$H(z)$, from either supernovae or BAO 
data, one can determine $w_{eff}^{(H)}(z)$ and $\Omega_{eff}^{(H)}$ relatively easily. On the other 
hand, measurements of $w_{eff}^{(A)}(z)$ and $\Omega_{eff}^{(A)}$ are less straightforward, because 
one has to determine $dH/dz$, or equivalently $dD_A/dz$ and $d^2D_A/dz^2$ (see Appendix 
\ref{app:a} for relation between these quantities). For this reason, the uncertainty of $\Theta$ 
is dominated by uncertainties of $w_{eff}^{(A)}(z)$ and $\Omega_{eff}^{(A)}$. Finally, coefficients 
${\mathcal F}_i$'s that present the evolution of equation of state of various constituents, are 
determined by fitting the deviation of $H(z)$ from the null hypothesis of a $\Lambda$CDM 
cosmology. However, as we argued in Sec. \ref{sec:frieman}, there are strong degeneracies 
between ${\mathcal F}_i$'s and $\gamma (z)$ which can be resolved only with using other types 
of data, in particular matter anisotropies, see Sec. \ref{sec:forquinmg} for more details.

As an example, we estimate the uncertainty of $\Theta$ for the Euclid mission. For the 
parametrization $w_{eff} (z) = w_p + w_a z /(1+z)$, according to the Euclid-Red 
Book~\cite{euclidred}, the standard deviation for these coefficients are expected to be 
$\sigma_{w_P} \sim 0.015$ and $\sigma_{w_a} \sim 0.15$ for Euclid data alone, and $\sigma_{w_P} 
\sim 0.007$ and $\sigma_{w_a} \sim 0.035$ for Euclid+Planck data. No forecast for the expected 
uncertainty of $dH/dz$ is yet available. For this reason, we simply use error propagation rules 
to determine a rough estimation for $\sigma_{dH/dz}$ from available forecasts. We approximate 
$dH/dz$ with its definition as a difference ratio: $dH/dz \approx \Delta H/\Delta z$, then we 
use the general uncertainty propagation rule to a function of $n$ variables $f (x_1, \ldots, 
x_n)$:
\be
\sigma_f^2 = \sum_{i,j = 1, \ldots, n} \frac{\partial f}{\partial x_i}\frac{\partial f}
{\partial x_j} C_{ij} \label{errorprop}
\ee
where $C_{ij}$ is the covariance matrix for random variables $x_1, \ldots, x_n$. Assuming 
$\sigma_H/H \sim 1\%$, negligible error for $z$, and ${\mathcal F}_i \sim 1$, the dominant source 
of error in $dH/dz$ from $w (z)$. Because the coefficients of derivatives with respect to these 
parameters in (\ref{errorprop}) is roughly of the order of one, we estimate $\sigma_{dH/dz} / (dH/dz) 
\sim 10-15\%$. Functions $A(z)$ and $B(z)$ are related to $dH/dz$, see (\ref{bzdef}), and when the 
uncertainties of $H$, density fractions $\Omega_i$'s and redshift $z$ are much smaller, 
$\sigma_{w_{eff}^{(A)}} / w_{eff}^{(A)}(z) \sim \sigma_{\Omega_{eff}^{(A)}} / \Omega_{eff}^{(A)} \sim 
\sigma_{\sigma_B} \sim \sigma_{dH/dz} \sim 10\%$ around optimal redshift of $z \sim 0.5$. 
Measurement precisions of ${\mathcal F}_i$'s also are of the order of precision of $dH/dz$, i.e. 
$\sigma_{{\mathcal F}_i} / {\mathcal F}_i \sim \sigma_{dH/dz} / (dH/dz)\sim 10-15\%$. 

Evidently, uncertainties obtained here are 
very rough estimations. The aim of these exercises is just to show what level of error we expect 
from near future surveys. A proper prediction needs detailed simulation of measurements and 
data analysing methods, instrumental effects, and systematic and statistical errors. They 
need a dedicated study that we leave to a future work.

Finally we want to make a remark about the redshift dependence of $w(z)$, which in the 
literature is usually parametrized~\cite{psobs}. In Appendix \ref{app:a} we show that for the 
same value of $w$ at two different redshifts, different parametrizations lead to very 
different evolution for $A(z)$. Inversely, if we estimate $w(z)$ from the measurement of 
$A(z)$, parametrization of $w(z)$ can lead to very different evolution for this function, 
despite employment of the same data for $A(z)$. Therefore, we must estimate $w$ at each 
redshift without parametrizing it. As for the estimation of uncertainties, for instance from 
the Fisher matrix, they can be determined from the set of $\{w(z),\gamma(z),z\}$ at every 
redshift bin rather than from a parametrization, see Appendix \ref{app:b}

\subsection{Discrimination between modified gravity and interacting quintessence 
models} \label{sec:forquinmg}
If we observe a non-zero $\Theta$, then we must use the power spectrum and growth rate of 
perturbations to investigate the nature and origin of deviation from a cosmological constant. 
The comparison between evolution equation of modified gravity and interacting quintessence 
models in Sec. \ref{sec:perturbint} showed that their interaction currents are very different, 
and thereby the evolution of matter anisotropies and dark energy density in these models are 
not the same. In fact, if we could decompose the interaction current to terms proportional to 
scalar metric perturbations and matter density fluctuations, it were possible to distinguish 
between these models. However, in practice measured quantities are matter power spectrum and 
its growth rate $\mathbf{f}(z,k)$ defined as:
\be
\mathbf{f}(z,k) \equiv \frac{d\ln D}{d\ln a} = \frac{\delta'_m}{{\mathcal H} \delta_m}, \quad 
D \equiv \frac{\delta_m (z,k)}{\delta_m (z=0,k)} \label{growthrate}
\ee
The function $\mathbf{f}(z,k)$ is usually extracted from the power spectrum using a 
model~\cite{fextect0,fextect1,fextect2}, for instance a power-law for the primordial spectrum 
including its modification by Kaiser effect~\cite{kaisereff,nonlinpert0,nonlinpert1,nonlinpert2} 
and redshift distortion due to the velocity dispersion~\cite{zdistort}. 

To obtain the evolution equation of $\mathbf{f}(z,k)$, we replace potentials $\psi$ and $\phi$ 
by expressions depending only on $\delta_m \equiv \delta \rho_m / \bar{\rho}_m$ and 
$\theta_m \equiv ik_jv_{(m)}^j$. Assuming a negligible anisotropic shear at 
$z \lesssim \mathcal {O} (1)$ which concerns galaxy surveys, scalar metric perturbations - 
gravitational potentials - $\psi$ and $\phi$ can be determined from Einstein equations 
(\ref{einst00})-(\ref{einstij}):\footnote{In this section for the sake of simplicity of notation 
we consider that ${\mathcal F}_i$'s factors for species are included in $w_i$'s, i.e. 
$(1+z)^{3\gamma_i} {\mathcal F}_i$ is redefined as $(1+z)^{3\gamma_i(z)}$ and $w_i$ is obtained 
from (\ref{gammade}) using this redefined $\gamma_i$. Therefore, for interacting dark energy 
models $w_m$ is nonzero and in general depends on redshift.}
\bea
\phi = \psi &=& \frac{4\pi G \bar{\rho}_m}{k^2} \bigg (\delta_m + 3 (1+w_m) \frac{{\mathcal H}
\theta_m}{k^2} \biggr ) + \Delta \psi \label{phipsisol} \\
\Delta \psi &=&  \frac{4\pi G}{k^2} \biggl (\delta\rho_\varphi - 3 {\mathcal H} \delta \varphi 
(\bar{\rho}_\varphi + \bar{P}_\varphi)^{\frac{1}{2}}\biggr ) \label{deltapsi} \\
\phi' &=& -\frac{4\pi G \bar{\rho}_m {\mathcal H}}{k^2} \biggl (\delta_m + 
(3 + \frac{k^2}{{\mathcal H}^2}) (1+w_m) \frac{{\mathcal H} \theta_m}{k^2} \biggr ) + 
\Delta \phi' \label{phip} \\
\Delta \phi' &=& -{\mathcal H}\Delta \psi + 4\pi G a^2 \delta \varphi (\bar{\rho}_\varphi + 
\bar{P}_\varphi)^{\frac{1}{2}}  \label{deltaphip}
\eea
Note that in (\ref{phipsisol}) and (\ref{phip}) we have separated terms which vanish for 
$\Lambda$CDM model and written them as $\Delta \psi$ and $\Delta \phi'$. As observations show 
that dark energy behaves very similar to a cosmological constant - at least for $z \lesssim 
\mathcal {O} (1)$, both these quantities are expected to be very small. It is why we write them 
as a variation of $\psi$ and $\phi'$. For future use it is also better to redefine them as 
followings:
\bea
&& \epsilon_0 \equiv \frac{\delta \rho_\varphi}{\bar{\rho}_m}, \quad \epsilon_1 \equiv 
\frac{{\mathcal H} (\bar{\rho}_\varphi + \bar{P}_\varphi)^{\frac{1}{2}} \delta \varphi}
{\bar{\rho}_m} \label{epsilo01def} \\
&& \Delta \psi = \frac{4\pi G \bar{\rho}_m}{k^2} (\epsilon_0 -3 \epsilon_1) 
\label{deltapsiredef} \\
&& \Delta \phi' = -\frac{4\pi G \bar{\rho}_m {\mathcal H}}{k^2} \biggl (\epsilon_0 - 
(3 + \frac{k^2}{{\mathcal H}^2}) \epsilon_1 \biggr ) \label{deltaphipredef}
\eea
After replacing $\phi'$ and $\psi$ in (\ref{deltarhoeq}) and (\ref{deltaveq}) with (\ref{phip}) 
and (\ref{phipsisol}) respectively, evolution equation of matter and velocity  perturbations 
can be written as:
\bea
&& \delta'_m + \frac{\bar{\rho}'_m}{\bar{\rho}_m} + 3 {\mathcal H} \biggl \{(1+ C^2_{sm}) 
\delta_m + (1+w_m) \frac{3\Omega_m {\mathcal H}^2}{2k^2} \biggl [\delta_m + 
(3 + \frac{k^2}{{\mathcal H}^2})(1+w_m) \frac {{\mathcal H}\theta_m}{k^2} \biggr ] + \nonumber \\
&& \hspace{2cm} (\epsilon_0 - (3 + \frac{k^2}{{\mathcal H}^2})\biggr \} + 
(1+w_m) \theta_m = \delta Q_{m0} \label{deltarhoeqeps} \\
&& \theta'_m + \frac{w'_m}{1+w_m} \theta_m + \frac{\bar{\rho}'_m}{\bar{\rho}_m} \theta_m + 
4 {\mathcal H} \theta_m - \frac{C_{sm}^2 k^2}{1+w_m} \delta_m - 3 \Omega_m {\mathcal H}^2 
\biggl (\delta_m + 3 (1+w_m) \frac{{\mathcal H} \theta_m }{k^2} + \epsilon_0 - 3\epsilon_1 
\biggr ) \nonumber \\
&& = ik_i \delta Q_{(m)}^i \label{deltaveqeps}
\eea
where $\delta Q_{m0}$ and $ik_i\delta Q_{(m)}^i$ are interaction currents and $\Omega_m \equiv 
8\pi G a^2 \bar{\rho}_m / 3 {\mathcal H}^2$. Moreover, in present 
and near future wide area surveys such as DES and Euclid the value of ${\mathcal H} / ck 
\ll 1$\footnote{Note that the speed of light $c=1$ is assumed in metric (\ref{metric}), and 
therefore it does not explicitly appear in our calculations.}. For instance, for Euclid 
${\mathcal H} / ck \lesssim 0.01$. Therefore, we can neglect terms proportional to 
${\mathcal H} / k$. Under these approximations, evolution equations of density and velocity 
become:

{\bf Modified gravity:}
\bea
&&\delta'_m + 3 {\mathcal H} (C_{sm}^2 - w_m) \delta_m + (1+w_m) \theta_m = \nonumber \\
&& \hspace{2cm} \frac {3\Omega_m (1-3w_m) {\mathcal C}_\varphi (\bar{\varphi})}{8\pi G} 
a {\mathcal H} \epsilon_1 + {\mathcal C} (\bar{\varphi}) \biggl (\frac {3\Omega_m (1-3w_m)}
{8\pi G}\biggr )^{\frac{1}{2}} \frac{\Omega_m (1+C_{s\varphi}^2)}
{2 \Omega_\varphi (1+w_\varphi)} a {\mathcal H} \epsilon_0 \label{deltarhoeqepsmgw} \\
&& \theta'_m + {\mathcal H} \theta_m - \frac{C_{sm}^2 k^2}{1+w_m} \delta_m - 
\frac{3\Omega_m}{2}{\mathcal H}^2 (\delta_m + \epsilon_0 - 3\epsilon_1) = 
- \frac{\sqrt {3} k^2 (1-3w_m) \Omega_m}{(8\pi G
(1+w_\varphi) \Omega_\varphi )^{\frac{1}{2}}} {\mathcal C} (\bar{\varphi}) \epsilon_1 
\label{deltaveqepsmgw}
\eea
{\bf Interacting quintessence:}
\bea
&& \delta'_m + 3 {\mathcal H} (C_{sm}^2 - w_m) \delta_m + (1+w_m) \theta_m = 
a A_{sm} \epsilon_0 \label{deltarhoeqepsqw} \\
&& \theta'_m + {\mathcal H} \theta_m - \frac{C_{sm}^2 k^2}{1+w_m} \delta_m - 
\frac{3\Omega_m}{2}{\mathcal H}^2 (\delta_m + \epsilon_0 - 3\epsilon_1) = 
-\frac{w_m}{1+w_m} (-L_m + A_{sm} \bar{\rho}_\varphi) a \theta_m \label{deltaveqepsqw}
\eea
Now that we have the evolution equations for $\delta_m$ and $\theta_m$, we can determine the 
evolution of growth rate. The procedure for calculating $d\mathbf{f}(z,k)/dz$ is 
straightforward. We replace $\theta_m$ in (\ref{deltaveqepsmgw}) and (\ref{deltaveqepsqw}) with 
its value obtained from (\ref{deltarhoeqepsmgw}) and (\ref{deltaveqepsqw}), respectively for 
modified gravity and interacting quintessence models. Then, we replace $\delta'_m$ with its 
value from equation (\ref{growthrate}). The final equation has the following general form:
\bea
&& \mathbf{f}' {\mathcal H} + \mathbf{f} ({\mathcal H}' + {\mathcal H}^2) + 
\mathbf{f}^2 {\mathcal H}^2 + 3 (C_{sm}^2 - w_m) ({\mathcal H}' + \mathbf{f} {\mathcal H}^2) + 
3 (C_{sm}^2 - w_m) {\mathcal H}^2 + \nonumber\\
&& \quad \quad \quad \frac{3}{2}~\Omega_m (1+w_m)^2 {\mathcal H}^2 + k^2 C_{sm}^2 + 
E_0 \mathbf{f} {\mathcal H} + E_1 k^2 + E_2 {\mathcal H} + E_3 {\mathcal H}^2 + E_4 = 0 
\label{growthratevol}
\eea
Coefficients $E_0,~E_1,~E_2,~E_3,~E_4$ depend on $z$ and $k$, and have the following values for 
the two models discussed here:

{\bf Modified gravity:}
\bea
E_0 & \equiv & {\mathcal C} (\bar{\varphi}) a (\bar{\rho}_\varphi + \bar{P}_\varphi)^
{\frac{1}{2}} \biggl (3 (C_{sm}^2 - w_m) + 1 - 3w_m \biggr ) \label {e0mg} \\
E_1 & \equiv & {\mathcal C} (\bar{\varphi}) ~\frac{(1+ w_m) (1 - 3w_m) \bar{\rho}_m \epsilon_1}
{{\mathcal H} (\bar{\varphi}) a (\bar{\rho}_\varphi + \bar{P}_\varphi)^{\frac{1}{2}}} 
\label {e1mg}\\
E_2 & \equiv & 3 (1+ w_m) (C_{sm}^2 - w_m) {\mathcal C} (\bar{\varphi}) a 
(\bar{\rho}_\varphi + \bar{P}_\varphi)^{\frac{1}{2}} - {\mathcal C}_\varphi (\bar{\varphi}) 
(1 - 3w_m) \frac{a \bar{\rho}_m \epsilon_1}{{\mathcal H} \delta_m} - \nonumber \\
&& {\mathcal C} (\bar{\varphi}) a (\bar{\rho}_\varphi + \bar{P}_\varphi)^{\frac{1}{2}} 
\biggl (\frac{(1 - 3w_m) (1+C_{s\varphi}^2) \bar{\rho}_m \epsilon_0}{2 (1+w_\varphi) 
\bar{\rho}_\varphi \delta_m} + 3 (C_{sm}^2 - w_m) \biggr ) \label {e2mg} \\
E_3 & \equiv & \frac{3\Omega_m (1+w_m)^2}{2} \biggl (\frac{\epsilon_0}{\delta_m} - 
\frac{3 \epsilon_1}{\delta_m}\biggr ) \label {e3mg} \\
E_4 & \equiv & -\frac{1}{\delta_m} \biggl ({\mathcal C}_\varphi (\bar{\varphi}) 
(1 - 3w_m) \frac{a \bar{\rho}_m \epsilon_1}{{\mathcal H}} + {\mathcal C} (\bar{\varphi}) a 
(\bar{\rho}_\varphi + \bar{P}_\varphi)^{\frac{1}{2}} \frac{(1 - 3w_m) (1+C_{s\varphi}^2) 
\bar{\rho}_m \epsilon_0}{2 (1+w_\varphi) \bar{\rho}_\varphi} \biggr)' + \nonumber \\
&& \biggl (3 {\mathcal C} (\bar{\varphi}) a (\bar{\rho}_\varphi + 
\bar{P}_\varphi)^{\frac{1}{2}} (C_{sm}^2 - w_m) \biggr )' - {\mathcal C} (\bar{\varphi}) a 
(1 - 3w_m) (\bar{\rho}_\varphi + \bar{P}_\varphi)^{\frac{1}{2}} \times \nonumber \\
&&\biggl [ {\mathcal C}_\varphi (\bar{\varphi}) 
(1 - 3w_m) \frac{a \bar{\rho}_m \epsilon_1}{{\mathcal H} \delta_m} + 
{\mathcal C} (\bar{\varphi}) a (\bar{\rho}_\varphi + \bar{P}_\varphi)^{\frac{1}{2}} 
\biggl (\frac{(1 - 3w_m) (1+C_{s\varphi}^2) \bar{\rho}_m \epsilon_0}{2 (1+w_\varphi) 
\bar{\rho}_\varphi \delta_m} + 3 (C_{sm}^2 - w_m) \biggr ) \biggr ] \nonumber \\
&& \label {e4mg}
\eea
{\bf Interacting quintessence:}
\bea
E_0 & \equiv & w_m a (-L_m + A_{sm} \bar{\rho}_\varphi) \label {e0quin} \\
E_1 & \equiv & 0 \label {e1quin} \\
E_2 & \equiv & 3 w_m a (C_{sm}^2 - w_m) (-L_m + A_{sm} \bar{\rho}_\varphi) + 
A_{sm} a \bar{\rho}_m (1+3w_m)\frac{\epsilon_0}{\delta_m}\label {e2quin} \\
E_3 & \equiv & \frac{3\Omega_m (1+w_m)^2}{2} \biggl (\frac{\epsilon_0}{\delta_m} - 
\frac{3 \epsilon_1}{\delta_m}\biggr ) \label {e3quin} \\
E_4 & \equiv & -A_{sm} a \bar{\rho}_m \biggl (\frac{\epsilon'_0}{\delta_m} + 
\frac{2a \epsilon_0}{\delta_m} (-L_m + A_{sm} \bar{\rho}_\varphi) \biggr ) \label {e4quin}
\eea
In the calculation of (\ref{growthratevol})-(\ref{e4quin}) we have neglected terms proportional 
to ${\mathcal H}/ck$. 

For $\Lambda$CDM model $E_i = 0,~i=0,\cdots,4$. For a non-interacting quintessence model 
all $E_i$ coefficients are zero except $E_3$. A notable difference between modified gravity and 
interacting quintessence models is the coefficient $E_1$ which is strictly zero for interacting 
dark energy models and nonzero for modified gravity that leaves an additional scale dependent 
signature on the evolution of matter anisotropies. The other explicitly scale dependent term is 
common for all models and is expected to be very small because it is proportional to the 
square of sound speed which is very small for cold matter. In addition, in contrast to the 
rest of $E_i$ coefficients, $E_1$ and $E_3$ are dimensionless. Evidently, the contribution 
of $E_1 k^2$ term with respect to other terms in equation (\ref{growthratevol}) increases for 
larger $k$, i.e. at short distances. But, the effect of nonlinearities, i.e. mode coupling also 
increases at large $k$, see e.g.~\cite {mgperturb}. They can imitate interactions and lead to 
misinterpretation of data. For this reason, it is suggested that observation of galaxy clusters 
is a good discriminator between dark energy models~\cite{mgcluster,mgcluster1}, because 
clusters are still close to linear regime, but have relatively large $k$.

Discriminating power of a survey can be estimated by the precision of $E_1$ and $E_3$ 
measurements. 
However, one expects some degeneracies when equation (\ref{growthratevol}) is fitted 
to determine $E_i$'s. Moreover, in galaxy surveys, $\mathbf {f}$ and $\mathbf {f}'$ (or more 
exactly $d\mathbf {f} / dz$) are determined from the measurement of power spectrum from 
galaxy distribution, and${\mathcal H}$ and ${\mathcal H}'$ from the BAO effect on the 
spectrum. Thus, these measurements are not completely independent. An independent measurement 
of ${\mathcal H}$ and ${\mathcal H}'$ e.g. using supernovae will help to reduce degeneracies 
and error propagation from measured quantities to the estimation of $E_i$'s. The relation 
between ${\mathcal H}'$ and $B(z)$ defined in (\ref{rhoderdl}) shows the logical connection 
of parametrization of homogeneous component - background cosmology - and evolution of 
fluctuations, specially in what concerns discrimination between dark energy models. In fact, 
anisotropies depend on the equation of state of matter, which in the context of interacting 
dark energy models, is modified by its interaction with dark energy. Thus, their independent 
measurements optimize their employment in distinguishing between various models.

Although apriori $d\mathbf {f} / dz$ can be determined directly from data by differentiating 
$\mathbf {f}$, usually due to shot noise the errors would be very large unless we extensively 
rebin the data. However, rebinning smears the redshift-dependence, which is the most important 
information for discriminating between models. Another approach is to solve equation 
(\ref{growthratevol}) analytically. It does not have an analytical solution for the general 
case, but as we show in Appendix \ref{app:d}, when $w_m$ and $C_{sm}^2$ are approximated by 
constant values, and cosmology is matter, radiation or cosmological constant dominated, i.e. 
up to desired precision only one component determines its evolution, an approximate solution can 
be found. At present epoch where 
matter and dark energy have comparable contributions, coefficients in (\ref{growthratevol}) 
even for $\Lambda$CDM vary with redshift. Nonetheless, their variation arrives very quickly 
to saturation. Therefore, the true solution is not very different from the approximate 
analytical one under the explained conditions, and it is possible to determine perturbations 
around the analytical solution by linearizing equation (\ref{growthratevol}), see 
Appendix \ref{app:d} for more details. 

A rough estimation of the uncertainties of $E_i$'s measured by Euclid can be performed in the 
same manner as what is presented in Sec. \ref {sec:forcc} for $\Theta$ and ${\mathcal F}_i$'s. 
It is expected that growth rate $\mathbf {f}$ can be reconstructed from Euclid+Planck data 
with an uncertainty $\sigma_{\mathbf {f}}/ \mathbf {f} \lesssim 3\%$~\cite{euclidred}. 
Considering equation (\ref {growthratevol}) and estimation of uncertainty of ${\mathcal H}'$ 
obtained in Sec. \ref{sec:forcc}, the uncertainty of $\sigma_{\mathbf {f}'}/ \mathbf {f}'$ must be 
$\sim 10\%-15\%$. This limits our ability to distinguish between a $\Lambda$CDM model where 
$E_i = w_m = 0$, and quintessence or interacting dark energy models where these quantities are 
not zero. Considering the linear equation obtained in Appendix \ref{app:d} from expansion of 
$\mathbf {f}$ around its solution for $\Lambda$CDM, the total uncertainty of deviation from this 
model is roughly the same as what is obtained for $\mathbf {f}'$, i.e. $\sim 10\%-15\%$. But, 
the uncertainty in the estimation of each $E_i$ is expected to be larger because of the 
degeneracy of these parameters. Evidently, determination of $\mathbf {f}$ and $\mathbf {f}'$ at 
multiple redshifts should help somehow reduce degeneracies and improve discrimination between 
models. More precise estimations as well as the estimation of the effect of nonlinearities 
and the optimal choice of scale range need detail simulation of surveys. We leave these tasks   
for future works. 

\subsection {Interpretation and comparison with other parametrizations} 
\label{sec:comare}
It would be useful to have a better insight on the physical meaning of the parameters defined 
in the previous section, and to compare them with what is used in the literature for 
parametrizing dark energy models.

We begin with $\epsilon_0$ and $\epsilon_1$ defined in (\ref{epsilo01def}). Their definitions 
show that the former depends only on dark energy density anisotropy and the latter only on the 
peculiar velocity of dark energy field, i.e. on its kinematics, see (\ref{uscalar}). They follow 
each other closely and approach zero when the field approaches its minimum value. However, 
their exponent close to the minimum depends on the interaction. Therefore, their measurements 
give us information about the potential and interactions of the scalar field. Moreover, the 
difference in the dependence of evolution equation of anisotropies and growth factor to 
$\epsilon_0$ and $\epsilon_1$ shows that only by separation of kinematics and dynamics of dark 
energy - scalar field - it would be possible to distinguish between modified gravity and other 
scalar field models.

The deviation of gravity potentials $\phi$ and $\psi$ from their value in $\Lambda$CDM 
$\Delta\psi$ is the quantity which can be measured directly from lensing data~\cite{lensing}. 
For this reason various authors have used $\Delta\psi$ to parametrize the deviation from 
$\Lambda$CDM~\cite{param0,param1,param2,param3}. However, equations (\ref{phipsisol}) and 
(\ref{deltapsi}) show that although $\Delta\psi \neq 0$ is by definition a signature of 
deviation from $\Lambda$CDM, in contrast to claims in the literature, it is not necessarily 
the signature of a modified gravity model because quintessence models, both interacting and 
non-interacting, also induce $\Delta\psi \neq 0$. This is also another manifestation of the 
difference between kinematics and dynamical effects of interacting dark energy models 
described above.

Because we have used Einstein frame for both quintessence and modified gravity models, in 
absence of an anisotropic shear $\phi = \psi$ even in non-$\Lambda$CDM models. In linear 
approximation gravitational lensing effect depends on the total potential 
$\Phi \equiv \phi + \psi$ (see e.g.~\cite{lensing} for a review). Therefore, in Einstein frame
\be
\Phi = 2\phi = 2\psi = \Phi_{\Lambda\text{CDM}} + 2\Delta\psi, \quad \quad 
\Phi_{\Lambda\text{CDM}} \equiv \frac{4\pi G \bar{\rho}_m}{k^2} \bigg (\delta_m + 3 (1+w_m) 
\frac{{\mathcal H}\theta_m}{k^2} \biggr ) \equiv \frac{4\pi G \bar{\rho}_m}{k^2 \Delta_m} 
\label{totpotent}
\ee
In the notation of~\cite{param0} $\Phi = 2\Sigma \Phi_{\Lambda\text{CDM}}$, thus: 
\be
\Sigma = 1 + \frac{\Delta\psi}{\Phi_{\Lambda\text{CDM}}} = \frac{\epsilon_0 - 3\epsilon_1}
{k^2 \Delta_m} \label{sigmalens}
\ee
The other quantity which affects the evolution of lensing and directly depends on cosmology is 
the growth factor of matter anisotropies which determines the evolution of $\Delta_m$ defined 
in (\ref{totpotent}). This quantity can be obtained from integration of growth rate $\mathbf{f}$ 
defined in (\ref{growthrate}) and is usually parametrized as $\Omega_m^{\gamma}$. For $\Lambda$CDM 
$\gamma \approx 0.55$~\cite{growthparam}. In this respect there is no difference 
between our formulation and what is used in the literature. Evidently, this simple 
parametrization cannot distinguish between various dark energy models. By contrast, the more 
sophisticated decomposition proposed in Sec. \ref{sec:forquinmg} is able to 
distinguish between quintessence and modified gravity. Note that in Jordan frame there are two 
other parameters: $\eta \equiv (\psi - \phi)/\phi$ and $Q = \phi/\phi_{\Lambda\text{CDM}}$. The 
parameter $\Sigma = Q (1+\eta/2)$, thus it is not independent. In Einstein frame $\eta = 1$ 
unless there is an anisotropic shear. At first sight it seems that there is less information 
in Einstein frame about modified gravity than in Jordan frame. However, one should notice that 
in Einstein frame the fundamental parameters are $\epsilon_0$ and $\epsilon_1$ and other 
quantities such as $\Delta\psi$ and $\mathbf{f}$ can be explained as a function of these 
parameters. Therefore, the amount of information in Einstein and Jordan frame about modified 
gravity - if it is what we call dark energy - is the same. The advantage of formulation in 
Einstein frame and definition of $\epsilon_0$ and $\epsilon_1$ is that they can be used for 
both major categories of models. Moreover, they have explicit physical interpretations that 
can be easily related to the underlying model of dark energy.

More recently based on an original work by C. Skordis~\cite{modgrskordis}, two 
groups~\cite{newparam,newparam0} have suggested new parametrizations which are basically 
only for discriminating modified gravity models from $\Lambda$CDM. Both groups use the following 
approximate description for the Einstein equation:
\be
G_{\mu\nu} = R_{\mu\nu} - \frac{1}{2} g_{\mu\nu} R = 8\pi G T_{\mu\nu} + U_{\mu\nu}
\label{einsteinskordis}
\ee
The tensor $U_{\mu\nu}$ is called {\it Energy-momentum tensor of dark energy}~\cite{modgrskordis}, 
and originally its definition has been for formulation of all modifications of the 
Einstein theory of gravity. In~\cite{newparam} this tensor is expanded with respect to 
potentials $\psi$ and $\phi$, and coefficients of this expansion are used for parametrizing 
the underlying modified gravity model.

Note that equation (\ref {einsteinskordis}) is at all scales an approximation because the right 
hand side is explicitly proportional to the Newton coupling constant. Considering $f(R)$ models 
which are the simplest modification of the Einstein theory of gravity, in contrast to 
(\ref{einsteinskordis}), the coupling to matter is modified in both frames, see equations 
(\ref{grmg0})-(\ref{grgmener}) for Jourdan frame, and the formulation of $f(R)$ model in 
Einstein frame in~\cite{frbean}. In fact, in Einstein frame the modification is explicit in the 
energy-momentum conservation equation. This means that if a deviation from $\Lambda$CDM is observed, 
it would be very difficult to verify the consistency of the model at short distances because the 
deviation of coupling from Newton constant $G$ is put by hand to zero. Moreover, this formulation 
and parametrization by definition does not help to detect interaction between dark energy and 
matter, because it depends only on the total variation of metric potentials. In addition, in this 
formulation $U_{\mu\nu}$ is assumed to be a conserved component, which as we discussed in 
Sec. \ref{sec:interact}, is not consistent because in contrast e.g. to perturbative quantum field 
theories, we never measure the {\it free component}. Furthermore, equation (\ref{einsteinskordis}) 
has exactly the same form for quintessence models, thus in this framework it is not possible to 
discriminate between this class and modified gravity models without knowing the underlying model in 
detail.

The formulation in~\cite{newparam0} uses a Lagrangian formalism with quadratic and higher order 
deviations from the Einstein theory of gravity. The energy-momentum tensor of dark energy $U_{\mu\nu}$ 
is obtained be using variational methods from this Lagrangian. It is a function of $g_{\mu\nu}$ or 
the set $\{g_{\mu\nu},~\varphi,~\partial_\mu \varphi\}$ when the dark-(energy) sector includes also a 
scalar field. Then, they use 3+1 spacetime decomposition, thus all coefficients of the above 
expansion depend only on time, and apply variational methods to determine perturbations 
$\delta U_{\mu\nu}$ around an arbitrary background. Their formulation is technically and theoretically 
interesting, specially for studying various modified gravity models, but there is neither a model 
independent parametrization for dark energy nor for observables.

\section{Outline} \label{sec:outline}
We have parametrized the interaction between dark energy and matter for modified gravity and 
interacting quintessence models as modifications of the evolution of matter and radiation background 
and perturbations densities, and the equation of state of dark energy. We have showed that when the 
interaction is ignored in the data analysis, the effective value of parameters are not the same if we 
calculate them from Friedman equation or from a function proportional to the derivative with respect 
to redshift of total mean energy density of the Universe. We have also defined a single quantity that 
evaluates the strength of the interaction. Its observational uncertainty can be used to estimate 
the discriminating power of a cosmological survey.

We have obtained a phenomenological description for the interaction current in the context of 
interacting quintessence models motivated by particle physics. Based on these results, we have 
suggested to distinguish between modified gravity and (interacting)-quintessence dark energy models of 
non-gravitational origin by the way they modify energy-momentum conservation equation. If the  
interaction current is proportional to the trace of the energy-momentum tensor of matter, we 
classify the model as {\it modified gravity}, otherwise, as {\it (interacting)-quintessence} and 
its variants, such as K-essence, quintom, cosmon, etc.

We have determined the modification of evolution equation of density and velocity perturbations 
in the context of modified gravity and interacting quintessence models discussed above, and 
used them to obtain a parametrized description of evolution equation of the growth factor that 
can be used for both these models as well as a simple $\Lambda$CDM model, which has been 
considered as the null hypothesis in our discussions. The difference between the value of these 
parameters can distinguish between aforementioned models. We have also obtained order of magnitude 
estimations for uncertainties on these quantities measured with the Euclid mission. A better 
forecast for these uncertainties needs simulations of the survey and the data analysis that we have 
left to future works.

\section*{Acknowledgment} The author would like to thank Luca Amendola and Ariel 
Sanchez for useful discussions about calculation techniques of cosmological parameters from 
LSS surveys data.

\appendix
\section{Properties of $A(z)$} \label{app:a}
\begin {center}
\begin {figure}
\begin {tabular}{lll}
\includegraphics[width=4cm,angle=-90]{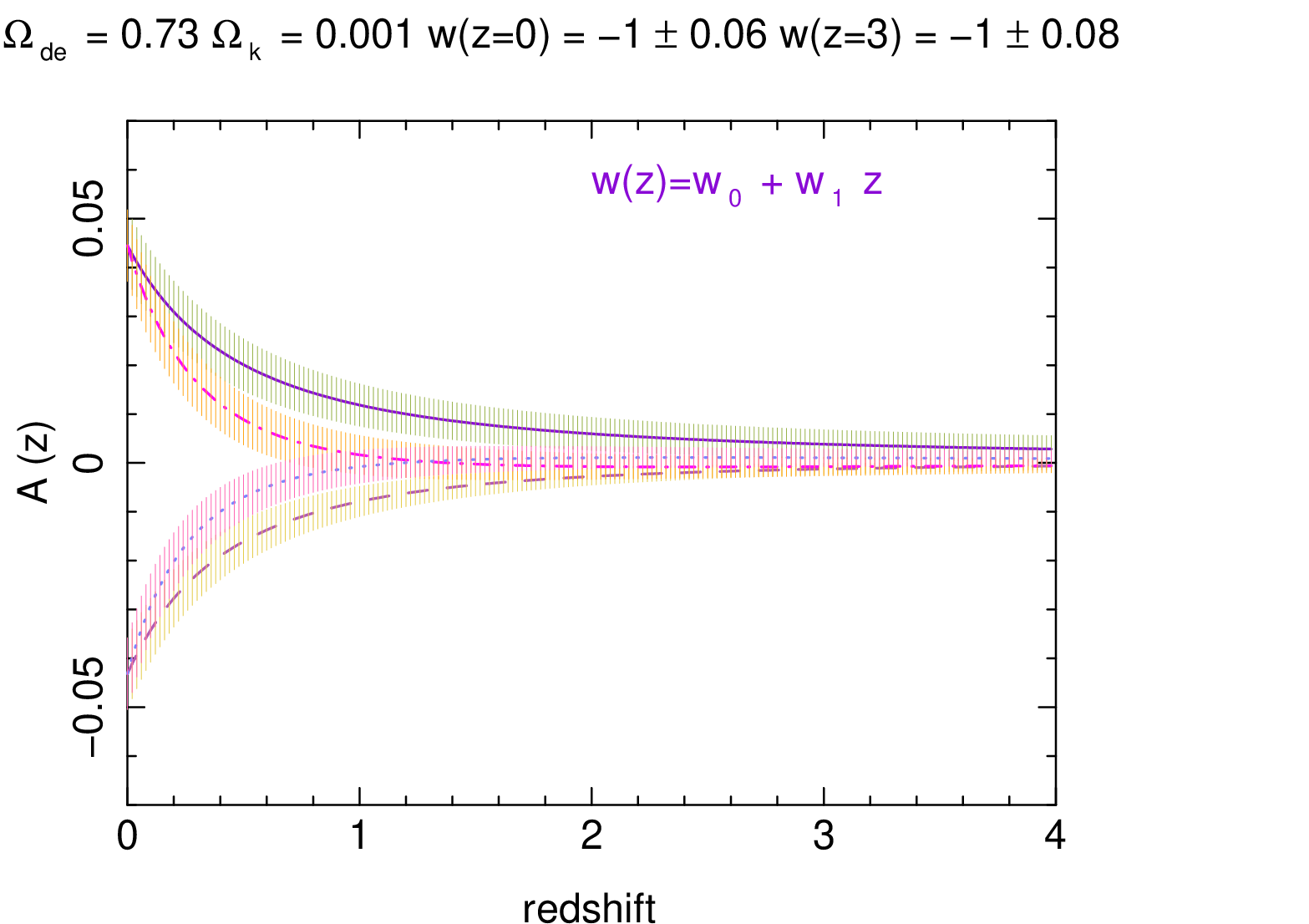} &\includegraphics[width=4cm,angle=-90]
{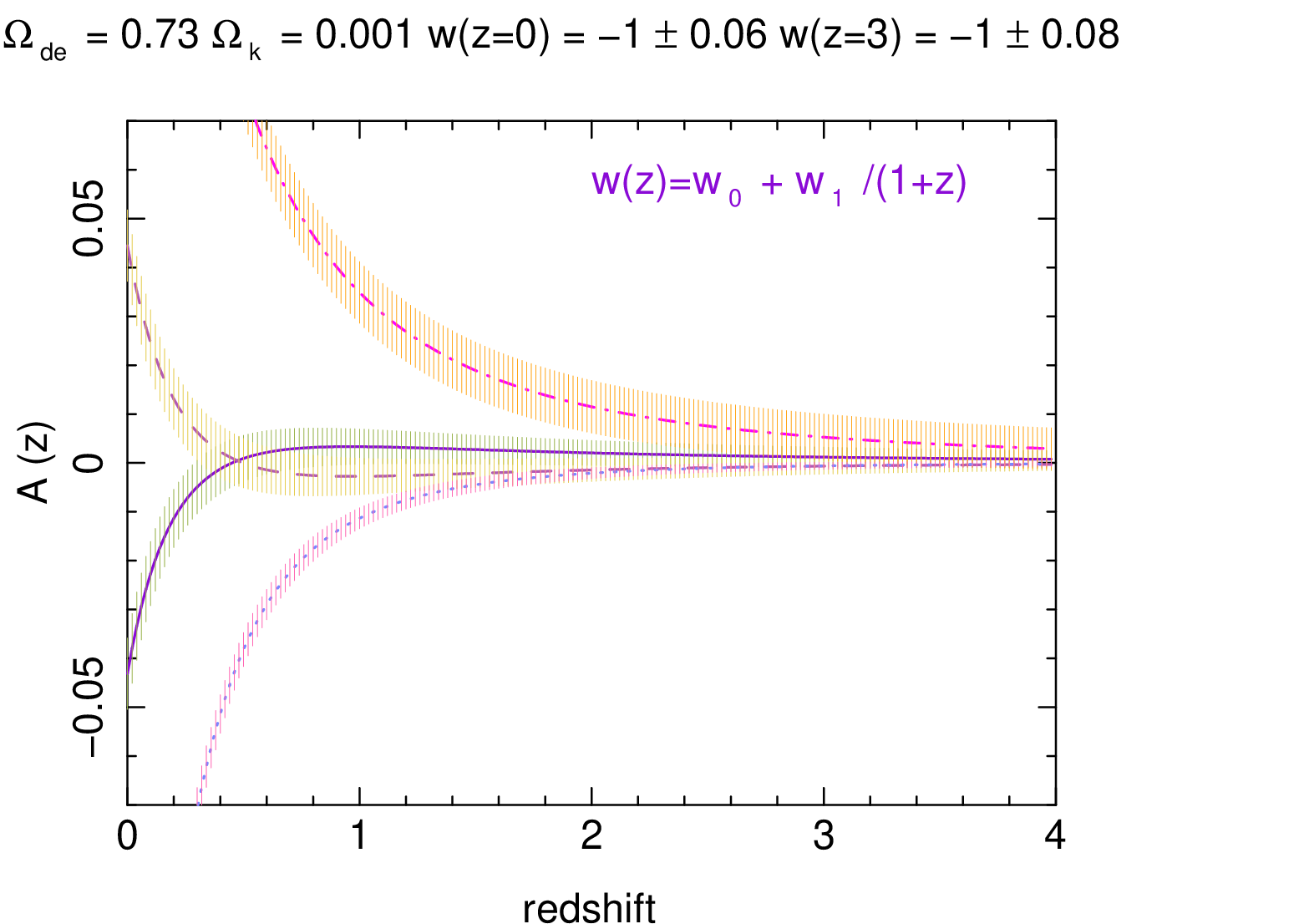} & \includegraphics[width=4cm,angle=-90]{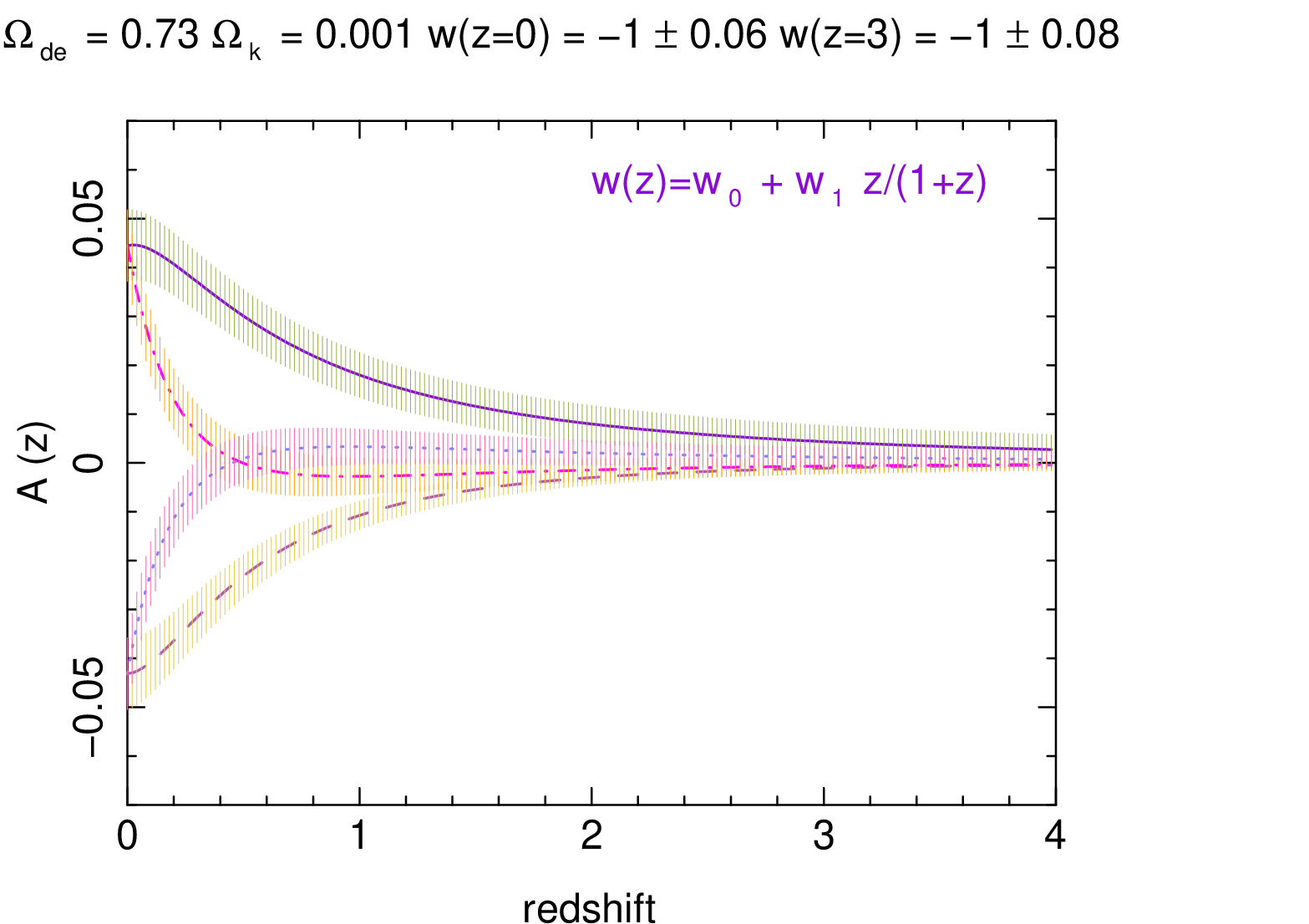} \\
\includegraphics[width=4cm,angle=-90]{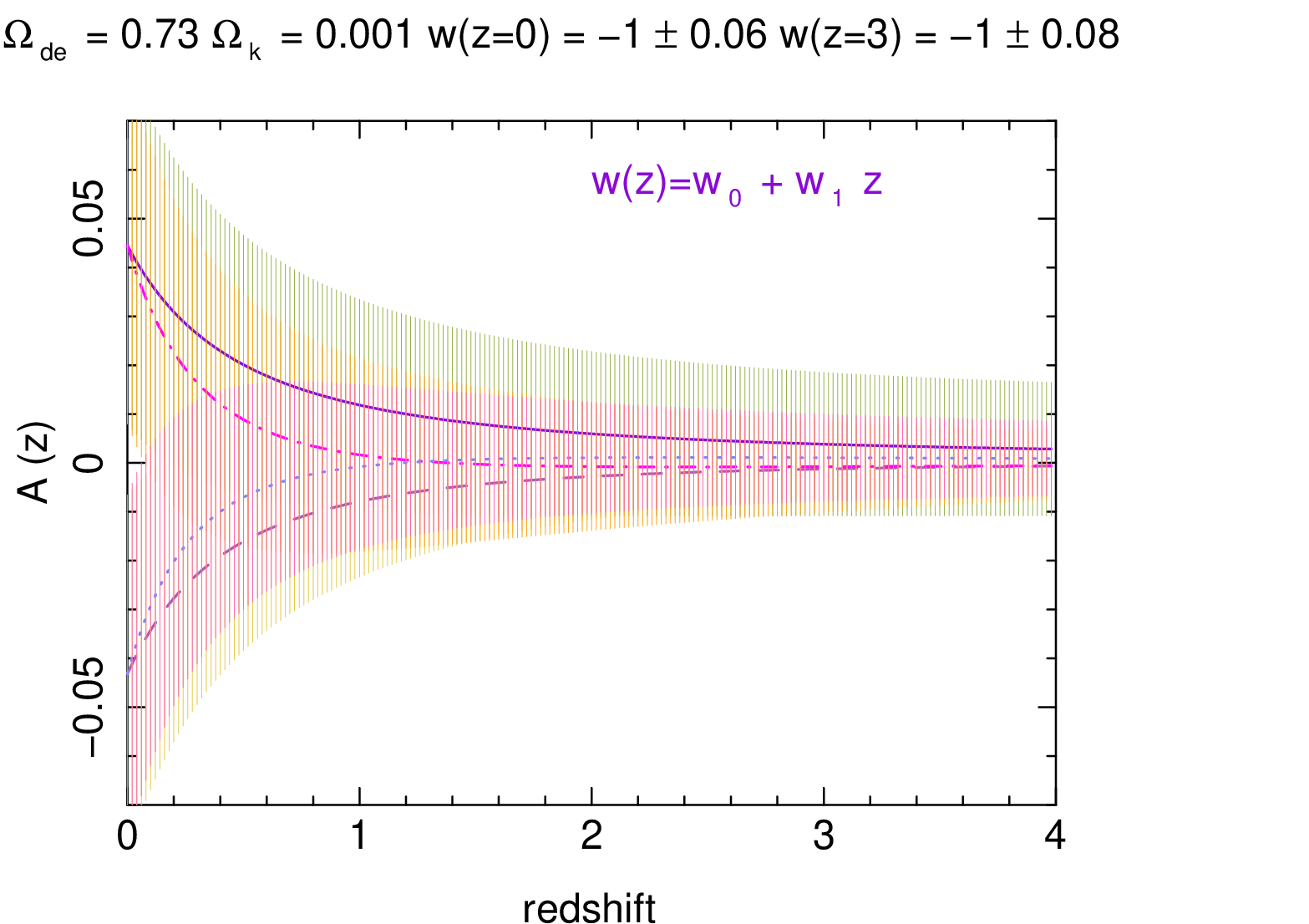} &\includegraphics[width=4cm,angle=-90]
{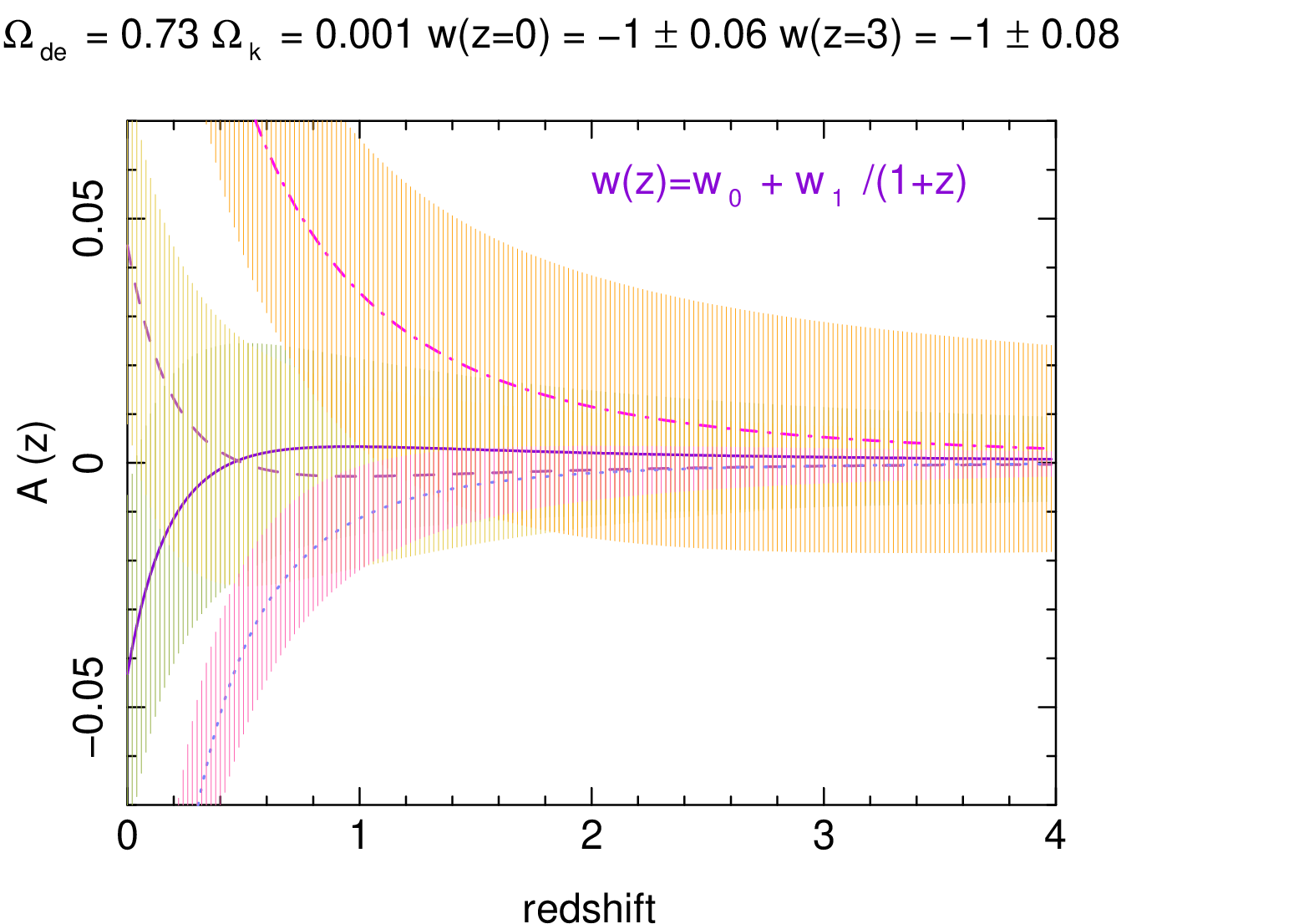} & \includegraphics[width=4cm,angle=-90]{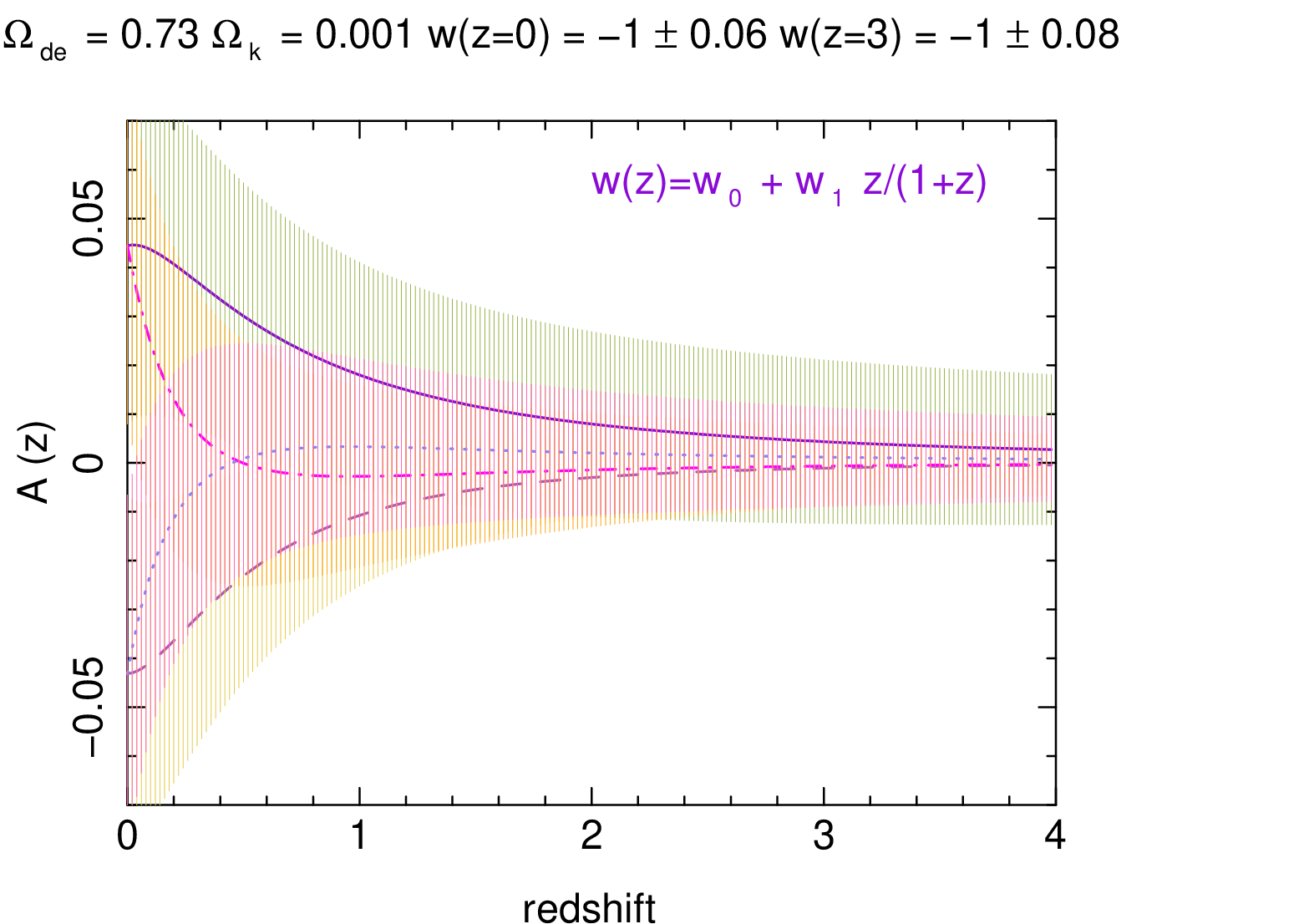}
\end {tabular}
\caption{$A(z)$ as a function of redshift. To see how well $A(z)$ can distinguish between various 
models and how systematic and statistical errors as well as parametrization affect the reconstructed 
model, we consider 3 parametrizations as written on the plot above. Note that parametrizations for 
the plot in the center and on the right are equivalent up to a redefinition of coefficients $w_0$ 
and $w_1$. We first consider a given value for $w(z)$ at $z=0$ and $z=3$, determine corresponding 
coefficients $w_{0i}$ and $w_{1i}$ where index $i$ is for initial. Then to simulate systematic errors 
we plot the following models: $w_0 = -1 + |w_{0i} +1|$, $w_1 = w_{1i}$ (dotted line), $w_0 = -1 - 
|w_{0i} +1|$, $w_1 = -w_{1i}$ (dot-dash), $w_0 = -1 + |w_{0i} +1|$, $w_1 = -w_{1i}$ (dashed) and 
$w_0=w_{0i}$ and $w_1=w_{1i}$ (full line). Colored vertical bars present statistical errors. The 
uncertainty of $az$ is $1\sigma_{A(z=0)} = 0.01$ (top row) and $1\sigma_{A(z=0)} = 0.05$ (bottom row) 
at $z=0$ and evolves with redshift as $\sigma_A(z) = \sigma_A(z= 0)(1+z)^2$. It seems to be 
possible to distinguish between normal and phantom dark energy models easily, if uncertainties are 
limited to few percents. Evidently, achieving such a precision is challenging even for space missions 
such as Euclid. \label{fig:wzparam}}
\end {figure}
\end {center}

One of the principle aims of LSS surveys is the measurement of Hubble constant $H(z)$, angular 
diameter distance $D_A$, and luminosity distance $D_L$, mainly by measuring Baryon Acoustic 
Oscillations (BAO) which play the role of a reference distance scale~\cite{bao}. The maximum 
effect of BAO on the power spectrum is at redshift $\sim 0.3$~\cite{bao}. 
However, as we mentioned in the Introduction a direct determination of $\gamma(z)$ from 
Hubble constant, $D_A$, or $D_L$ when $z \rightarrow 0$ is not possible. In fact, using equation 
(\ref{friedmannoint}) and the definition of angular diameter distance, it is easy to see that:
\be
\ln \biggl [\biggl (\frac{d}{dz}((1+z) D_A) \biggr)^{-1} - \Omega_m (1+z)^3 - \Omega_h 
(1+z)^4 - \Omega_K (1+z)^2\biggr ] = \ln \Omega_{de} + 3\gamma (z) \log (1+z) \label{loghubble}
\ee
At small redshifts the last term on the r.h.s. of (\ref{loghubble}) which contains $\gamma (z)$ 
approaches zero, and the effect of the latter becomes negligibly small. Now, consider the following 
quantities:
\bea
H^2 (z) &=& \frac{8\pi G}{3} \rho (z) \label {dlhh} \\
B(z) &\equiv& \frac{1}{3 (1+z)^2 \rho_0} \frac {d\rho}{dz} = \frac{2 H(z)}{3 H_0^2(1+z)^2}~
\frac{dH}{dz} = \frac{2{\mathcal H}(z)}{3(1+z){\mathcal H}_0} \biggl (\frac{(1+z) 
d{\mathcal H}}{dz} + {\mathcal H} \biggr ) \label{bzdef}\\
A(z) &\equiv& B(z) - \Omega_m - \frac{4}{3}\Omega_h (1+z) - \frac{2\Omega_K}{3 (1+z)} 
\nonumber \\
&=& \Omega_{de}\biggl(\gamma + (1+z) \ln (1+z) \frac{d\gamma}{dz}\biggr)(1+z)^{3 (\gamma - 1)}
= \Omega_{de}(w(z) + 1)(1+z)^{3 (\gamma - 1)} \label{azdef}
\eea
where $H (z) = \dot{a} /a$ is the expansion rate of the Universe and $\rho (z)$ is the total 
density at redshift $z$. It is clear that $A(z)$ is proportional to the deviation of dark energy 
from a cosmological constant at any redshift including $z = 0$. In addition, its sign determines 
whether dark energy has normal or phantom-like equation of state at a given redshift. It can be 
shown~\cite{houriaz} that when $dw/dz \ll 3w(z)(w(z)+1)/(1+z)$, the sign of $dA/dz$ is opposite 
to the sign of $w(z)+1$. This condition is satisfied at low redshifts - see examples of models 
in Fig. \ref{fig:wzparam}. It means that $A(z)$ is a concave or convex function of redshift, 
respectively for positive or negative $w(z)+1$. Observations show that the contribution of 
$\Omega_k$ and $\Omega_h$ at low redshifts is much smaller than the uncertainty of $\Omega_m$. The 
function $dA/dz$ does not depend on $\Omega_m$. Thus, the uncertainty on the value of $\Omega_m$ 
can shift the value of $A(z)$ but it does not change its slope and its shape i.e. its concavity 
or convexity will be preserved.

The function $B(z)$ can be easily related to directly measurable quantities:
\bea
&& B(z) \equiv \frac{1}{3 (1+z)^2 \rho_0} \frac {d\rho}{dz} = 
\frac{\frac{2}{1+z}(\frac{dD_l}{dz} - \frac {D_l}{1+z}) - 
\frac{d^2D_l}{dz^2}}{\frac{3}{2} (\frac{dD_l}{dz} - \frac {D_l}{1+z})^3}
\label {rhoderdl} \\
&& D_l = (1+z) H_0 \int_0^z \frac {dz}{H (z)} \label {dlh} \\
\eea
or equivalently with respect to normalized angular distance:
\bea
&& B(z) = \frac{-(2 \frac{dD_A}{dz} + (1+z) \frac{d^2D_A}{dz^2})}
{\frac{2}{3(1+z)^2} (D_A + (1+z)\frac {dD_A}{dz})^3}
\label {rhoderda} \\
&& D_A = \frac{H_0}{1+z}\int_0^z \frac {dz}{H (z)} = \frac{D_l}{(1+z)^2} \label {dah}
\eea
Note that these equations are written for a flat universe, but can be easily extended to the cases 
where $\Omega_k \neq 0$.

\section {About Fisher matrix for equation of state of dark energy} \label{app:b}
Fisher matrix evaluates the sensitivity - information content - of a measured quantity to variables 
and parameters that define the underlying model~\cite{fisher}. Under special conditions, e.g. 
Gaussianity of distributions, Fisher matrix can be related to the covariance matrix of measurements. 
In LSS surveys the main measured quantity is the power spectrum of matter density anisotropies. 
Application of Fisher matrix to CMB~\cite{fishercmb} and galaxy 
surveys~\cite{fextect0,fextect1,fextect2} is well studied and widely used. In what concerns the 
measurement of dark energy density, its variation, and its equation of state from galaxy surveys, 
one has to extract $H(z)$ and $D_A(z)$ either from BAO~\cite{baoobs,baoobs1,baoobs2} or by fitting 
the complete power spectrum~\cite{psobs}. Fisher matrix for 2-dimensional power spectrum is 
determined by Seo \& Eisenstein~\cite{fextect0,fextect1,fextect2} with $H(z)$ 
and $D_A(z)$ as parameters. A transformation from these quantities to coefficients of a parametrized 
equation of state, for instance $w (z) = w_0 + w_a z/(1+z)$ allow to determine the covariant matrix 
for the measurement of $w_0$ and $w_a$~\cite{psobs}.

Although apriori the value of these quantities can be determined at any redshift, in practice 
the limited volume and deepness of surveys allow to determine the power spectrum at the average 
redshift of the survey or for some bins of redshift in the case of large deep surveys. In the latter 
case, the estimation of $w(z)$ as a function of redshift depends strongly on its parametrization. 
Fig. \ref{fig:wzparam} shows the plot of $A(z)$ for examples in which $w$ is measured at two 
redshifts. It is evident that this quantity and thereby the underlying dark energy models depend 
strongly on the parametrization of $w$, notably when systematic and statistical errors are added.

Simpson and Peacock~\cite{growthrate} use $\{w_0, w_a, \Omega_\Lambda, \Omega_k, \Omega_mh^2, 
\Omega_bh^2, n_s, A_s, \beta, \gamma', \sigma_p\}$ as independent parameters for estimating 
cosmological parameters from the measurement of the galaxy power spectrum. Here 
$w_a\equiv -dw/dz$, $\beta (z) \equiv f(z) / b(z)$ where $\mathbf {f}(z)$ is the growth rate of 
scalar fluctuations and $b(z)$ is the linear bias, and $\gamma'$ is the parameter that define an 
approximate parametrization for $\mathbf{f}(z) \approx \Omega_m^{\gamma'}(z)$ for 
$\Lambda$CDM~\cite{growthparam}. It can be also shown that in what concerns the determination of the 
Fisher matrix for dark energy, $w(z)$ and $dw/dz$ alone lead to a singularity~\footnote{For the sake 
of simplicity in the discussion of Fisher matrix here, we neglect other cosmological parameters, i.e 
we assume that dark energy parameters can be factorized from other quantities. In practice, one has 
to consider a single matrix Fisher matrix containing all parameters. Thus there would be one single 
covariant matrix that includes correlation of all uncertainties.}. 

In place of parametrizing $w(z)$, we suggest to use $w(z)$, $\gamma (z)$ and $z$ to determine the 
Fisher matrix for dark energy parameters. It can be easily shown that Fisher matrix becomes 
singular if the first two quantities are considered~\cite{psobs}, because $w(z)$ and $\gamma (z)$ 
are not independent - if one knows $w(z)$, then $\gamma (z)$ can be determined from (\ref{gammade}). 
This problem does not arise when $w$ is parametrized because expansion parameters are explicitly 
independent. The relationship of $w(z)$ and $\gamma(z)$ is very similar to the relation between 
$H(z)$ and $D_A(z)$. Fisher matrix for $\{H(z),D_A(z),z\}$ set of parameters is 
calculated in~\cite{psobs}. Using this formulation, a parameter transformation gives the Fisher 
matrix for $\{w(z),\gamma (z),z\}$. Relation between Fisher matrices with 2 sets of parameters 
${p_i}$ and ${q_m}$ is~\cite{fishercmb}:
\be
\bar{F}_{ij} = \sum_{mn} \frac{\partial q_m}{\partial p_i} F_{mn} \frac{\partial q_n}
{\partial p_j} \label{fishertrans}
\ee
For the parameter-sets discussed above, the components of the Jacobian matrix are:
\bea
\frac {\partial H(z)}{\partial w(z)} &=& \frac{3H_0^2 \Omega_{de}}{2H(z)} (1+z)^{3\gamma (z)} 
\label{hwjocob} \\
\frac{\partial H(z)}{\partial \gamma(z)} &=& \frac{3H_0^2 \Omega_{de}}{2H(z)} 
(1+z)^{3\gamma (z)}\ln (1+z) \label{hgammajocob} \\
\frac{\partial D_A (z)}{\partial w(z)} &=& - \frac{1}{H^2 (z)}\frac{\partial H(z)} 
{\partial w(z)} \label{dawjocob} \\
\frac{\partial D_A (z)}{\partial \gamma (z)} &=& -\frac{1}{H^2 (z)}\frac{\partial H(z)}
{\partial \gamma (z)} \label{dagammajocob} \\
\frac {\partial H(z)}{\partial z} &=& \frac{H_0^2}{2H(z)} \biggl (3\Omega_m (1+z)^2 + 
2 \Omega_k (1+z) + \frac{1+w(z)}{1+z}\biggr ) \label{hzjocob} \\
\frac{\partial D_A (z)}{\partial z} &=& -\frac{1}{1+z}\biggl (D_A (z) + \frac{1}{H(z)}\biggr ) 
\label{dazjocob}
\eea
Alternatively, one of $w(z)$ or $\gamma (z)$ parameters can be replaced by $A (z) = 
\Omega_{de}(w(z) + 1)(1+z)^{3 (\gamma - 1)}$. In fact, it is preferable to replace $\gamma (z)$ with 
$A(z)$, because at low redshifts the $\gamma(z)$-dependent term has very small effect on the 
evolution $H(z)$ and $D_A$. By contrast, the deviation of $A(z)$ from its value in $\Lambda$CDM 
model is maximized for $z \rightarrow 0$, see Fig. \ref{fig:wzparam}.

\section {Fluid description of a scalar field} \label{app:c}
Energy momentum tensor of a scalar field is:
\be
T^{\mu\nu}_\varphi = -\frac{1}{2} g^{\mu\nu} g^{\rho\sigma} \partial_\rho \varphi 
\partial_\sigma \varphi + g^{\mu\nu} V (\varphi) + \partial^\mu \varphi \partial^\nu \varphi 
\label{enmomscalar}
\ee
Using definition (\ref{tfluid}) of a perfect fluid, the density and pressure are defined as:
\be
\rho_\varphi \equiv u_\mu u_\nu T^{\mu\nu}_\varphi = \frac{1}{2} \partial^\mu \varphi 
\partial_\mu \varphi + V (\varphi), \quad \quad P_\varphi \equiv \frac{1}{2} \partial^\mu 
\varphi \partial_\mu \varphi - V (\varphi) \label{densscalar}
\ee
$u^\mu$ is the velocity vector and $u^\mu u_\mu = 1$. It is easy to verify that with above 
definitions for $\rho_\varphi$ and $P_\varphi$:
\be
u_\mu = \frac{\partial_\mu \varphi}{(\rho_\varphi + P_\varphi)^{\frac{1}{2}}} \label{uscalar}
\ee

\section {Solution of evolution equation of growth rate} \label{app:d}
For $\Lambda$CDM cosmology, $E_i=0,~i=0,\ldots,4$. We also consider $w_m = C^2_{sm} = 0$. In 
this case after dividing equation (\ref{growthratevol}) by ${\mathcal H}^2$, the evolution 
equation of growth rate becomes:
\be
\frac{\mathbf{f}'}{\mathcal {H}} + \mathbf{f} (\frac{{\mathcal H}'}{\mathcal {H}^2} + 1) + 
\mathbf{f}^2 + \frac{3}{2}~\Omega_m = 0 
\label{grwothtarelcdm}
\ee
After changing the variable from $\eta$ to $\ln a$, this equation changes to:
\be
\frac{d\mathbf{f}}{dx} + (\frac{x''}{x'^2} + 1) \mathbf{f} + \mathbf{f}^2 + 
\frac{3}{2}~\Omega_m = 0, \quad \quad x \equiv \ln \frac {a(\eta)}{a_0(\eta)} 
\label{grwothtarelcdma}
\ee
By integrating the Friedman equation for flat $\Lambda$CDM one obtains: 
\bea
{\mathcal H} &=& \frac{d\ln (\frac{a}{a_0})}{d\eta} = x'= \frac {{\mathcal H}_0 a}{a_0} 
\sqrt{\Omega_m(a_0) (\frac{a_0^3}{a^3}) + \Omega_\Lambda} \label {mathh} \\
{\mathcal E} \equiv \frac{{\mathcal H}'}{\mathcal {H}^2} &=& \frac {x''}{x'^2} = 
\frac{\Omega_\Lambda - \Omega_m (a_0)\frac{a_0^3}{a^3}}{\Omega_\Lambda + 
\Omega_m (a_0) \frac{a_0^3}{a^3}} = \frac{\Omega_\Lambda - \Omega_m (a_0) e^{-3x}}
{\Omega_\Lambda + \Omega_m (a_0) e^{-3x}} \label{hph2}
\eea
For $z=0$, ${\mathcal E} = -1$ and for $z \rightarrow \infty$, ${\mathcal E} = 
\Omega_\Lambda(a) - \Omega_m (a)$. To be able to solve (\ref{grwothtarelcdma}) analytically we 
must assume ${\mathcal E}$ is a constant. This is a good approximation if we are interested 
only on a small range of redshifts. Under this assumption, the solution of 
(\ref{grwothtarelcdma}) can be obtained by integration:
\be
\mathbf{f}_{\Lambda CDM} (z) \approx \frac{-({\mathcal E} + 1 - \frac{\alpha_1}{2}) + 
({\mathcal E} + 1 + \frac{\alpha_1}{2})(1+z)^{\alpha_1}}{1 - (1+z)^{\alpha_1}}, \quad \quad 
\alpha_1 = \sqrt {({\mathcal E} + 1)^2 - 6 \Omega_m} \label{fsol}
\ee
For $-\sqrt{6\Omega_m} - 1 < {\mathcal E} < \sqrt{6\Omega_m} - 1$, $\alpha_1$ is imaginary and 
according to this approximation solution $\mathbf{f}(z)$ has an oscillating component. A simple 
attempt to make (\ref{fsol}) more precise is to take into account that ${\mathcal E}$ depends on 
redshift.

To obtain an approximate solution for interacting dark energy models parametrized by coefficients 
$E_i=0,~i=0,\ldots,4$ in (\ref{growthratevol}), under the assumption that these corrections are small, 
we can linearize this equation around $\mathbf{f}_{\Lambda CDM}$. Note that in general it is expected that 
in interacting dark energy models $w_m$ and $C^2_{sm}$ are not zero. Therefore, we add also their 
contribution to the linearized model:
\bea
&& \mathbf{f} = \mathbf{f}_{\Lambda CDM} + \Delta \mathbf{f} \label{growthratedecomp} \\
&& \Delta \mathbf{f}' + \biggl [\frac{{\mathcal H}'}{{\mathcal H}} + E_0 + {\mathcal H} 
\biggl (1 + 3 (C_{sm}^2 - w_m) + 2~\mathbf{f}_{\Lambda CDM} \biggr) \biggr ] \Delta \mathbf{f} + 
3 (C_{sm}^2 - w_m)\frac{{\mathcal H}'}{{\mathcal H}} + \nonumber \\
&& \quad \quad \quad 3 {\mathcal H} \biggl (C_{sm}^2 - w_m + \frac{\Omega_m}{2} w_m 
(2 + w_m)\biggr ) + (C_{sm}^2 + E_1) \frac{k^2}{{\mathcal H}} + E_2 + E_3 {\mathcal H} + 
\frac{E_4}{{\mathcal H}} = 0 \label{deltagrowthrate}
\eea
Solution of this linearized equation is straightforward and can be formally written as the 
following:
\bea
\Delta \mathbf{f} (z) &=& \frac{{\mathcal H}}{(1 + z)(1 + 3 (C_{sm}^2 - w_m))}~\exp \biggl [\int 
\frac{dz}{1+z} \biggl (\frac{E_0}{{\mathcal H}} + 2~\mathbf{f}_{\Lambda CDM} \biggr ) \biggr ] 
\times \nonumber \\
&& \biggl \{1 + \int dz~\frac{(1 + z)(1 + 3 (C_{sm}^2 - w_m))}{(1+z){\mathcal H}^2} 
~\exp \biggl [-\int \frac{dz}{1+z} \biggl (\frac{E_0}{{\mathcal H}} + 2~\mathbf{f}_{\Lambda CDM} 
\biggr ) \biggr ] \nonumber \\
&& \biggl [3 (C_{sm}^2 - w_m) \frac{{\mathcal H}'}{{\mathcal H}} + 3 {\mathcal H} 
\biggl ((C_{sm}^2 - w_m) + \frac{\Omega_m}{2}w_m (2+w_m) \biggr ) + \frac{k^2}{{\mathcal H}} 
(C_{sm}^2 + E_1) + E_2 + \nonumber \\
&& E_3 {\mathcal H} + \frac{E_4}{{\mathcal H}} \biggr ]\biggr \} 
\label{deltagrowsssol}
\eea
Determination of integrals in (\ref{deltagrowsssol}) needs details of redshift dependence of 
coefficients $E_i$'s which is model dependent. Nonetheless, they depend on the scalar field which must 
vary very slowly with redshift. Therefore, at zero order, they can be considered as constant. 
Although even with this simplification it is difficult to determine (\ref{deltagrowsssol}) 
analytically, a numerical determination allow to write it as an expansion with respect to $E_i$ 
coefficient. This expansion would be suitable for compression with data and determination of $E_i$.
\end{document}